\documentclass[11pt]{article}
\usepackage[T1]{fontenc}
\usepackage{lmodern,amsmath,amssymb,amsthm,booktabs,array,tikz}
\usepackage[a4paper,left=25mm,right=25mm,top=23mm,bottom=25mm]{geometry}
\usepackage[hidelinks]{hyperref}
\hypersetup{pdftitle={Potentially planar degree sequences: characterization and complexity},
  pdfauthor={Zhe Zhang and Renwen Tian}}
\newtheorem{theorem}{Theorem}[section]
\newtheorem{lemma}[theorem]{Lemma}
\newtheorem{corollary}[theorem]{Corollary}
\newtheorem{proposition}[theorem]{Proposition}
\newcommand{\vol}{\operatorname{vol}}
\newcommand{\problem}{\textnormal{\textsc{Planar Degree Realization}}}
\definecolor{accent}{HTML}{175F73}
\definecolor{retained}{HTML}{A4382B}
\tikzset{v/.style={circle,fill=black,inner sep=1.65pt},pin/.style={circle,draw=accent,fill=white,inner sep=2pt},keep/.style={line width=1pt,retained},guide/.style={accent,dashed,thick}}
\title{Potentially planar degree sequences:\\ characterization and complexity}
\author{Zhe Zhang\\[2pt]
  {\small Independent Researcher}\\
  {\small Shanghai, China}\\
  {\small\href{mailto:zz940521@gmail.com}{\nolinkurl{zz940521@gmail.com}}}
  \and
  Renwen Tian\\[2pt]
  {\small School of Mathematical Sciences}\\
  {\small Beijing Normal University, Beijing, China}\\
  {\small\href{mailto:renwen.tian@mail.bnu.edu.cn}{\nolinkurl{renwen.tian@mail.bnu.edu.cn}}}}
\date{}
\makeatletter
\newcommand{\tablecaption}{\def\@captype{table}\caption}
\newcommand{\figurecaption}{\def\@captype{figure}\caption}
\renewcommand{\maketitle}{%
{\centering
{\LARGE\bfseries \@title\par}
\medskip{\large\begin{tabular}[t]{c}\@author\end{tabular}\par}
}}
\makeatother
\begin{document}
\maketitle
\begingroup
\setlength{\leftmargini}{1em}
\begin{abstract}
In 1963, Hakimi asked which sequences of nonnegative integers admit a simple planar realization. We answer this question with two exact criteria, an interval rule that tracks residual degrees along a disk boundary and a nonlinear finite-difference test in the degrees alone. However, recognition is strongly NP-complete even for graphic sequences of length $n$ with minimum degree at least three and sum $6n-12$, and the sorted degree sequences of two $13$-vertex triangulations have a graphic midpoint with no planar realization, ruling out any characterization by simultaneous affine inequalities in the sorted entries even at sum $66$. On an $L$-bit input, a deterministic $O(L)+2^{O(\sqrt{n})}$-time search finds a realization or rules one out. For positive graphic sequences of length $n$ with $\ell$ ones, we classify all exceptions at the first two even sums above the sparse bound of Bar-Noy et al., $4n-2-2\ell$ and $4n-2\ell$. This gives linear-time recognition on a word RAM throughout the extended range.
\end{abstract}
\endgroup

\section{Introduction}
Hakimi's 1963 problem asks which \emph{graphic} lists (degree sequences of simple graphs) admit a planar realization~\cite[Appendix~IV, Problem~35]{bm}. We characterize these \emph{potentially planar}, or \emph{planaric}, lists and prove that recognizing them is strongly NP-complete. The octahedron realizes $(4^6)$ (exponents denote multiplicities). At the same sum, the three universal vertices of $(5^3,3^3)$ force a $K_{3,3}$ with the three cubic vertices. Appending one two makes the list planaric (Figure~\ref{fig:compact-six}). In \problem\ the input is an explicit binary list $d=(d_1,\dots,d_n)$ of nonnegative integers. A realization on $[n]$ must give vertex $i$ degree $d_i$ and may have several components or isolated vertices.

B\"ottger and Harders~\cite{bh} exhibited a nonplanaric list satisfying Hakimi's necessary conditions~\cite{hakimi63}. Chv\'atal~\cite[Problem~24]{ckk} asked for necessary and sufficient conditions, citing the Erd\H{o}s--Gallai criterion for graphicness. Schmeichel and Hakimi~\cite{sh} later bounded sums of the largest degrees. Our interval rule keeps residual degrees in boundary order (Theorem~\ref{th:compact-interval}). A thirteen-vertex midpoint rules out simultaneous affine inequalities in the sorted degrees (Proposition~\ref{prop:compact-midpoint}), but counting sphere embeddings gives a nonlinear test (Theorem~\ref{th:compact-arithmetic}). At the same triangulation sum $6n-12$, we encode \textsc{3-Partition}.

\begin{theorem}\label{th:hard}
\problem\ is strongly NP-complete even for $n\ge4$, minimum degree at least three and degree sum $6n-12$. Every output of the reduction is graphic and has a $3$-connected realization. Every yes-instance produced has a stacked triangulation realization.
\end{theorem}

At this sum every planar realization is a triangulation. Fanelli and Zaffiri~\cite{fz} studied such lists under connectivity or adjacency constraints. The reduction in Section~\ref{sec:encoding} uses triangular faces to encode prescribed adjacencies. Polynomial-time recognition on graphic inputs would imply $\mathrm{P}=\mathrm{NP}$. Polynomially checkable certificates for both answers, a good characterization in the sense of Edmonds~\cite{edmonds}, would imply $\mathrm{NP}=\mathrm{coNP}$. For arbitrary inputs a weighted separator bounds the boundary data needed to construct a realization.

\begin{theorem}\label{th:alg}
On a deterministic multitape Turing machine, \problem\ can be decided in $O(L)+2^{O(\sqrt n)}$ time, where $L$ is the bit length of the delimited binary list. Within the same bound the algorithm returns the edges of a realization on the input labels whenever one exists.
\end{theorem}
Listing the $2^{\Theta(\sqrt n)}$ planaric degree multisets of order $n$ takes $2^{O(\sqrt n)}$ total time (Corollary~\ref{cor:compact-listing}).

How sparse must a graphic list be to guarantee a planar realization? For positive graphic $n$-term lists with $\ell$ ones, Bar-Noy et al.~\cite{sp:mfcs,bn} proved that a sum of at most $4n-4-2\ell$ suffices. At $4n-2-2\ell$ they left open whether an obstruction can have a leaf or more than six vertices. Only $(4^5,2)$ fails there (Theorem~\ref{sp:thm:first}), and Theorem~\ref{sp:thm:next} classifies the next layer, $4n-2\ell$.

An appended two raises this boundary by four and the sum by two. For a graphic list $h$ as in Theorem~\ref{th:hard} with $p=|h|\ge8$, the boundary is reached after $p-6$ twos and the resulting list is planaric (Corollary~\ref{sp:cor:eight}). Section~\ref{sec:open-budget} asks how large $T(h)$, the minimum needed for planarity, can be.

\section{Criteria for general degree lists}\label{sec:general}
\subsection{An ordered interval criterion}\label{sec:compact-interval}
Each move updates residual degrees, as in the Havel--Hakimi rule~\cite{sp:havel,sp:hakimi}. The choice is nondeterministic and respects the order of the positive residual degrees on a disk boundary.
Discard the zeros and start with an empty row. Remove one unused occurrence of a degree $d$ and choose a consecutive block of $k\le d$ row entries. For $k\ge2$, every strict interior entry must equal one. Replace the block $a_1,\ldots,a_k$ by $(a_1-1,d-k,a_k-1)$. For $k=1$, subtract one from the chosen entry and insert $d-1$ on either side. For $k=0$, insert $d$ at any gap, including either end. Erase zeros after every move.

\begin{theorem}\label{th:compact-interval}
A list of nonnegative integers has a simple planar realization if and only if some sequence of these moves consumes all its positive entries and ends with an empty row.
\end{theorem}
\begin{proof}
Draw the partial graph in a disk with the vertices still needing neighbours on its boundary in row order. Join a new vertex outside the disk to the block by a noncrossing fan. The strict interior vertices receive their last neighbour, so enlarging the disk around the fan gives the new row. An accepting run realizes the list once the isolates are restored.

Conversely, complete a planar realization on the positive entries to a plane triangulation on the same vertices (orders below three are immediate). Colour the original edges red. Follow a canonical ordering from a base edge~\cite{dfpp}, recording along the outer path the positive counts of red neighbours still to come. Insert the two base vertices first, counting their edge if red. Each new vertex covers an interval of the outer path. Every strict interior vertex still in the row leaves the path now, so its sole remaining red neighbour is the new vertex. The red predecessors thus form a legal block, with any unchosen endpoints retained on their respective sides. After the last vertex the row is empty.
\end{proof}

\begin{figure}[b]
\centering
\begin{tikzpicture}[x=0.9cm,y=0.9cm,line cap=round,line join=round]
\foreach \panel/\offset in {0/-2.6,1/2.6}{
\begin{scope}[xshift=\offset cm]
\coordinate (a) at (-1.8,0); \coordinate (b) at (1.8,0); \coordinate (c) at (0,1.65);
\coordinate (u) at (0,2.64);
\coordinate (x) at (0,.22); \coordinate (y) at (.72,.78); \coordinate (z) at (-.72,.78);
\draw (a)--(u)--(b) (c)--(u) (a)--(x)--(b) (b)--(y)--(c);
\draw[guide] (a)--(b)--(c)--cycle;
\ifnum\panel=0
  \draw (a)--(y);
  \draw[preaction={draw=white,line width=2.6pt}] (c)--(x);
  \node[left,font=\scriptsize] at (a){5}; \node[right,font=\scriptsize] at (b){5};
  \node[left,font=\scriptsize] at (c){5};
  \node at (0,-.45) {\small $(5^3,3^3)$};
\else
  \draw[keep] (c)--(z)--(a) (x)--(y);
  \node[pin] at (z){};
  \node at (0,-.45) {\small $(5^3,3^3,2)$};
\fi
\foreach \j in {a,b,c,u,x,y}{\node[v] at (\j){};}
\end{scope}}
\end{tikzpicture}
\caption{The nine solid edges on the left form $K_{3,3}$ and the dashed triangle joins the degree-five vertices. The bold edge and the path through the open circle replace the two crossing edges.}\label{fig:compact-six}
\end{figure}
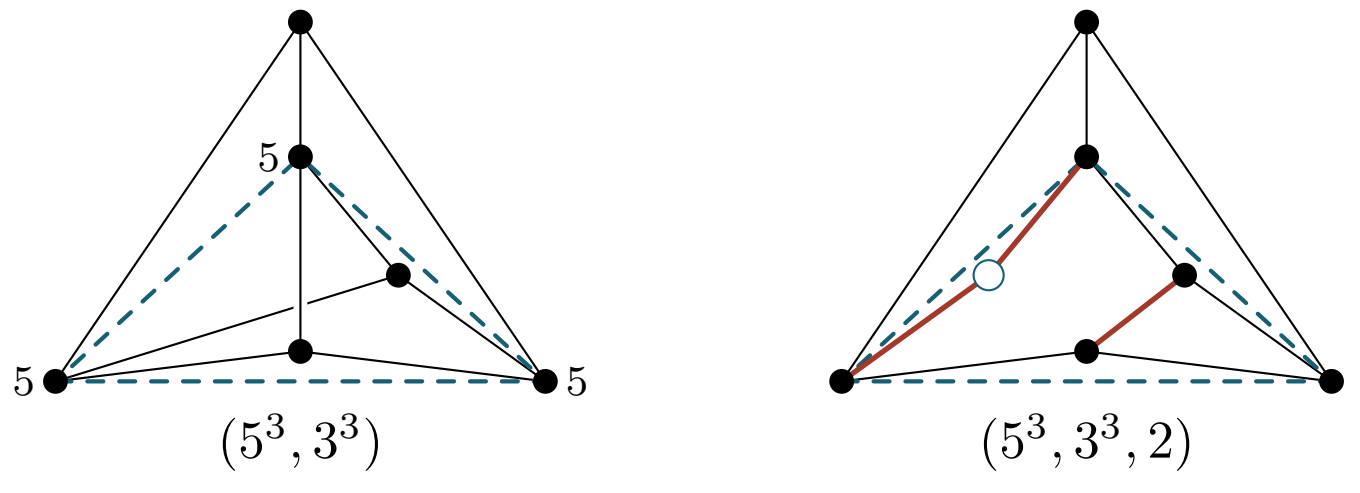

An accepting run is a polynomial certificate for the NP-complete recognition problem of Theorem~\ref{th:hard}. Free row permutations would admit $(5^3,3^3)$. Join two degree-five vertices, bring them together to attach each cubic vertex to both, then join the last vertex to all five (Figure~\ref{fig:compact-six}).

\subsection{A midpoint obstruction on thirteen vertices}\label{sec:compact-midpoint}
The interval rule maintains a boundary order. The Erd\H{o}s--Gallai test~\cite{sp:eg} for graphicness uses only the sorted degrees. At even sum its minima expand into simultaneous affine inequalities.
\begin{proposition}\label{prop:compact-midpoint}
For the sorted lists
\[
 h_j=(12,7+j,7-j,5^5,3^5),\qquad j=0,1,2,
\]
$h_0$ and $h_2$ have stacked planar realizations, and their midpoint $h_1$ is graphic and has no planar realization. Thus no system of affine inequalities in the sorted entries characterizes planar realizability even among graphic lists of length thirteen and sum $66$.
\end{proposition}
\begin{proof}
Each list has sum $66=6\cdot13-12$. Every planar realization is a triangulation. Deleting its universal vertex leaves a triangulated $12$-gon. Coning its boundary reverses the operation.

For $h_0$, label a heptagon cyclically from $0$ to $6$, draw diagonals $35,36,26,16$, and attach an ear to each boundary edge except $12$ and $56$. For $h_2$, use the fan at $0$ and omit the ears on $23$ and $45$. Coning the two polygons in Figure~\ref{fig:compact-midpoint} gives the stated degrees. Successive ear deletions reduce each cone to a tetrahedron, so both triangulations are stacked.

\begin{figure}[htbp]\centering
\begin{tikzpicture}[scale=.91,line join=round,
  nucleus/.style={fill=black,circle,inner sep=1.25pt},
  ear/.style={fill=white,draw=black,circle,inner sep=1.65pt}]
\foreach \shift/\diagonals/\attached in
  {0/{3/5,3/6,2/6,1/6}/{0,2,3,4,6},
   5.2/{0/2,0/3,0/4,0/5}/{0,1,3,5,6}} {
  \begin{scope}[xshift=\shift cm]
  \foreach \i in {0,...,6}
    \coordinate (v\i) at ({90-360*\i/7}:1.38);
  \draw[thick] (v0)--(v1)--(v2)--(v3)--(v4)--(v5)--(v6)--cycle;
  \foreach \i/\j in \diagonals \draw[thin] (v\i)--(v\j);
  \foreach \i in \attached {
    \pgfmathtruncatemacro{\j}{mod(\i+1,7)}
    \coordinate (e\i) at ({90-360*(\i+.5)/7}:1.82);
    \draw[thick] (v\i)--(e\i)--(v\j);
    \node[ear] at (e\i) {};
  }
  \foreach \i in {0,...,6} {
    \node[nucleus] at (v\i) {};
    \node[font=\scriptsize,fill=white,inner sep=.5pt]
       at ({90-360*\i/7}:1.68) {$\i$};
  }
  \end{scope}
}
\end{tikzpicture}
\caption{The polygons for $h_0$ and $h_2$ before the degree-twelve vertex is added. Their degrees are $(6^2,4^5,2^5)$ and $(8,4^6,2^5)$ respectively. The numbered vertices form the heptagon and the open vertices are its five added ears.}\label{fig:compact-midpoint}
\end{figure}
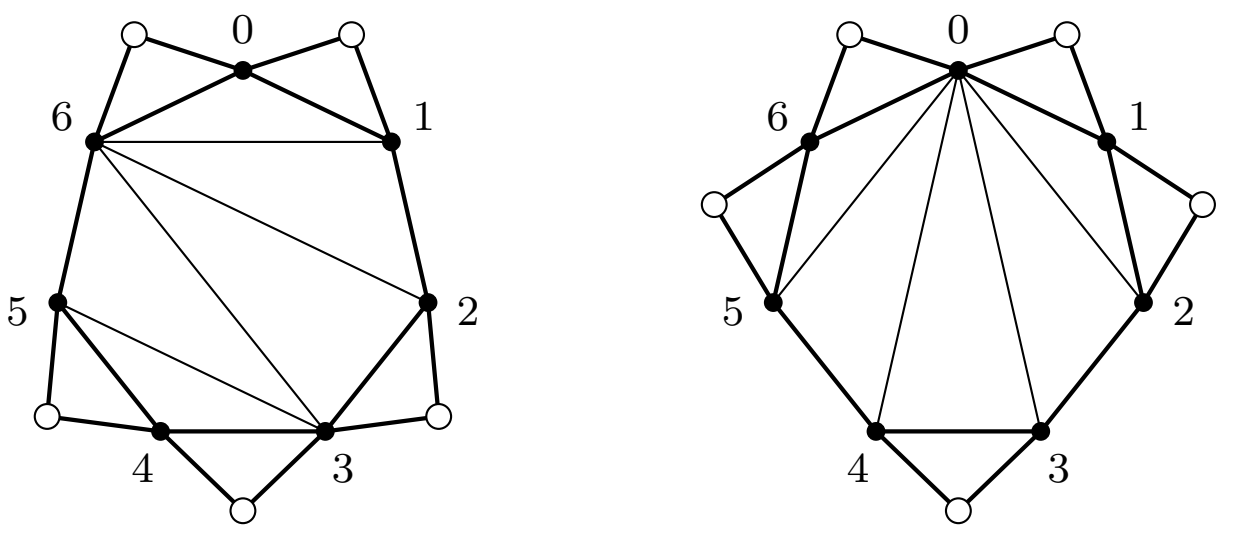

The polygon for $h_1$ would have degrees $(7,5,4^5,2^5)$. Remove its five pairwise nonadjacent ears. Call the two boundary edges of the remaining heptagon $P$ that carried no ear \emph{bare}. If $v$ meets $b_v$ bare edges, its former degree was $d_P(v)+2-b_v$. At the vertex $A$ of former degree seven, $d_P(A)=5+b_A\le6$. If $b_A=1$, then $P$ is a fan at $A$. The bare edge at $A$ ends at a fan vertex of degree two whose former degree was at most three, below the required four.

If $b_A=0$, the five neighbours of $A$ form its link path $U$. The sole vertex $X$ outside this fan forms an ear on a link edge $yz$. Both boundary edges at $X$ carried ears since its former degree was at least four. For $v\in\{y,z\}$, only link edges other than $yz$ can be bare. Thus $b_v\le d_U(v)-1$ and $d_P(v)=2+d_U(v)$. Both $y$ and $z$ formerly had degree at least five, although the target allows only one such vertex besides $A$.

In the cone over the right polygon, vertices $0$ and $1$ have degrees nine and five. Replacing the diagonal $03$ by the absent edge $13$ gives a simple realization of $h_1$. Every affine inequality satisfied by the two endpoints holds at their midpoint.
\end{proof}

Thirteen is the first order with a missing integral midpoint at sum $6n-12$. Order three is immediate. Using \texttt{plantri}~\cite{plantri}, we checked all pairs of distinct sorted triangulation degree lists at orders four through thirteen. The $947$ lists at order thirteen have five distinct missing midpoints.

\subsection{An arithmetic criterion in the degrees}\label{sec:compact-arithmetic}
Counting embeddings by their number of faces gives a nonlinear test in the degrees alone. It distinguishes the missing midpoint of Proposition~\ref{prop:compact-midpoint} from its two endpoints.

Discard zeros, accepting the empty list. For the remaining $n$ entries reject a degree at least $n$ or an odd sum. Write $2m=\sum_i d_i$. If $m\le n-1$, construct a forest by leaf removal. While a degree exceeds one, remove a one and decrement a largest degree. The sum bound guarantees a one and persists after each step. Pair the remaining ones and reverse the steps by attaching leaves. For $n\ge3$, reject $m>3n-6$.

Let $B$ be the oriented incidence matrix of $K_n$ with one row deleted. For integer $q\ge1$, let $X_i$ be $q\times q$ matrices of independent commuting variables, $\partial_i$ their matrices of entrywise partial derivatives, and $C$ diagonal with $c_{ij}=\operatorname{tr}(X_iX_j)$ for $i<j$. 
\begin{theorem}\label{th:compact-arithmetic}
Let $d$ have $n$ positive integer entries, even sum $2m$, and $\max_i d_i<n$. Suppose $n\le m\le3n-6$ and put $f=m-n+2$. Then $d$ has a simple planar realization if and only if
\[
 \sum_{q=1}^{f}(-1)^{f-q}\binom fq H_d(q)>0,\qquad
 H_d(q)=\left.\left(\prod_i\operatorname{tr}(\partial_i^{\,d_i})\right)
 \det\!\begin{pmatrix}I+C&CB^{\mathsf T}\\-B&0\end{pmatrix}\right|_{X=0}.
\]
\end{theorem}
Extracting the sphere term gives this algebraic existence criterion (Appendix~\ref{app:arithmetic-proof}). The separate face search in Theorem~\ref{th:alg} constructs a witness in $O(L)+2^{O(\sqrt n)}$ time.

\section{Exact recognition at sparse degree sums}\label{sp:sec:results}
One vertex of degree two suffices to make the midpoint $h_1$ of Proposition~\ref{prop:compact-midpoint} planaric. Cone the right polygon of Figure~\ref{fig:compact-midpoint} with apex $u$. Delete the edge from $u$ to vertex $0$ and draw a new two-edge path from $u$ to vertex $1$ alongside their retained edge. This realizes $(h_1,2)$, so the least numbers of twos needed for $h_0,h_1,h_2$ are $0,1,0$, respectively.

To count the twos needed in general, remove the zeros and write $d=(h_1,\ldots,h_p,2^t,1^\ell)$ with $h_i\ge3$.
Put $H=\sum_i h_i$, $n=p+t+\ell$ and $S=H+2t+\ell$. When $p=0$, every graphic list is planaric, since every component of any realization is a path or a cycle. Assume $p\ge1$ and $H+\ell$ is even. Let $T(h,\ell)$ be the smallest $t$ for which $d$ is planaric. We write $T(h)=T(h,0)$.

When $\ell>H$, attach $h_i$ separate leaves to vertex $i$ and pair the remaining leaves. This gives $T(h,\ell)=0$. Suppose henceforth that $\ell\le H$. Set $e=(H-\ell)/2$. Consider the planar multigraphs $M$ on the $p$ core vertices with exactly $e$ edges and $\deg_M(i)\le h_i$. Loops and parallel edges are allowed, and a loop contributes two to its vertex degree. Let $s(M)$ count distinct adjacent pairs of different vertices and let $\lambda(M)$ count loops with multiplicity.

\begin{lemma}\label{lem:compact-support}
The minimum number of degree-two vertices is
\[
 T(h,\ell)=e-\max_M\bigl(s(M)-\lambda(M)\bigr).
\]
The list $(h,2^t,1^\ell)$ is planaric exactly when $t\ge T(h,\ell)$.
\end{lemma}
\begin{proof}
The maximum is over a finite, nonempty set. Choose integers $0\le r_i\le h_i$ summing to $2e$. Pair the odd demands by disjoint edges and fill the remaining even demands with loops. For any candidate $M$, attach $h_i-\deg_M(i)$ leaves to vertex $i$. Keep one edge from each nonloop parallel class and subdivide each additional copy once. Subdivide each loop twice. The result is a simple planar graph with the required degrees using $e-s(M)+\lambda(M)$ twos. Further twos can be placed along a single edge.

Conversely, take a simple planar realization, discard components consisting entirely of twos, and suppress the paths through the remaining twos. In the resulting plane multigraph, each nonloop parallel class needed at least one subdivision per extra copy, and each loop needed at least two. If two leaves form an isolated edge $uv$, some core edge $xy$ must exist: with $b$ leaf--leaf edges the core has $e+b$ edges. Replace $uv,xy$ by pendant edges $xu,yv$, also when $x=y$. The leaves can be drawn at their new endpoints, and deleting a core edge cannot increase the required subdivisions. Repeat until all leaves are pendant. Removing them leaves a candidate $M$ whose subdivision cost is at most the original $t$.
\end{proof}

For a list with no ones or twos and $H=6p-12$, zero cost requires $s(M)-\lambda(M)=3p-6$. Since $M$ has exactly $3p-6$ edges, each must join a distinct pair. Its support is then a triangulation. Theorem~\ref{th:hard} concerns exactly this equality case.

For the sparse layers write $S=4n-2\ell-2j$. Substitution gives $e-t=2p-j$, so for $\ell\le H$ Lemma~\ref{lem:compact-support} reads
\begin{equation}\label{eq:compact-support}
 d\text{ is planaric}\quad\Longleftrightarrow\quad
 \max_M\bigl(s(M)-\lambda(M)\bigr)\ \ge\ 2p-j .
\end{equation}
Bar-Noy et al.~\cite[Theorem~2]{bn} prove that every positive graphic list with $j\ge2$ is planaric. For $p\ge1$, the next two layers automatically satisfy $\ell\le H$, since $H-\ell=4p+2t-2j$. Their targets in~\eqref{eq:compact-support} are $2p-1$ and $2p$. The following results classify precisely which graphic lists miss them.

\begin{theorem}\label{sp:thm:first}
A positive graphic list with $S\le4n-2-2\ell$ is planaric if and only if $d\ne(4^5,2)$, up to order.
\end{theorem}
The exceptional list was identified by Schmeichel and Hakimi~\cite[Theorem~4]{sh}. This answers the two questions of Bar-Noy et al.~\cite[Section~4.4]{sp:mfcs}, \cite[Section~7]{bn} about larger leaf-free obstructions and obstructions with a leaf at this sum.

\begin{theorem}\label{sp:thm:next}
A positive graphic list with $S=4n-2\ell$ has no planar realization exactly in the following cases, up to order:
\begin{enumerate}
\item $p=5$.
\item $p=6$ and exactly three of the $h_i$ are three.
\item $d$ is $(\ell+5,4^4,3,1^\ell)$, $(\ell+6,4^5,2,1^\ell)$, or $(\ell+6,5^2,3^4,1^\ell)$ for some $\ell\ge0$.
\item $d$ is $(5^5,3,2^2)$, $(4^7)$ or $(6^3,3^4,2)$.
\end{enumerate}
\end{theorem}

If $t\ge3$, only the first two cases remain. The graphic exception $(5^5,3,2^2)$ shows that this simplification already fails at $t=2$. Appendix~\ref{sp:app:proofs} proves both boundary theorems.

\begin{corollary}\label{sp:cor:eight}
On $S=4n-2\ell$, every positive graphic list with $p\ge8$ is planaric. For every $\ell$ there is an exception with $p=7$.
\end{corollary}
\begin{proof}
The exceptional list $(\ell+6,5^2,3^4,1^\ell)$ is graphic: add a vertex adjacent to all six vertices of $K_{2,4}$ and attach $\ell$ leaves to it.
\end{proof}
Through $S=4n-2\ell$, recognition takes deterministic linear time on a word RAM. After checking the degree bound and parity, a degree histogram and its prefix sums suffice for both the Erd\H{o}s--Gallai test~\cite{sp:kiraly} and the exceptions above.

Let $h$ be graphic with $p\ge8$, minimum degree at least three and $H=6p-12$. Subdividing an edge $p-6$ times gives a graphic list at $S=4n$. Thus Corollary~\ref{sp:cor:eight} gives $T(h)\le p-6$, improving the bound $p-4$ from Bar-Noy et al.~\cite{bn}. On this class, deciding $T(h)\le r$ is strongly NP-complete for every fixed integer $r\ge1$ and for $r=p/29-1$ when $29\mid p$ (Theorem~\ref{th:compact-fixed-twos}). Binary search with Theorem~\ref{th:alg} finds $T(h)$ and a witness in $2^{O(\sqrt p)}$ time after input reading, using $O(\log p)$ calls on lists of length at most $2p-6$.
\begin{proposition}\label{prop:compact-spectrum}
For each $p\ge4$, every integer $m$ from $0$ to $\lfloor(p-4)/2\rfloor$
occurs as $T(h)$ for a graphic list $h$ of length $p$, minimum degree at
least three and sum $6p-12$. In fact, with $q=2m+4$ and $k=p-q$, one may take
\[
 h=\bigl((p-1)^2,q-1,4^k,3^{q-3}\bigr).
\]
\end{proposition}
The construction and the matching lower bound are given in Appendix~\ref{app:consequences}.

\section{Encoding prescribed edges by degrees}\label{sec:encoding}
For a graphic list $h$ of length $n\ge4$, minimum degree at least three and sum $6n-12$, $T(h)=0$ exactly when $h$ is realized by a \emph{sphere triangulation}, a simple graph embedded on the sphere with triangular faces. The reduction for Theorem~\ref{th:hard} encodes prescribed edges by \emph{stacking} a vertex in a triangular face and joining it to the three corners. A triangulation built from $K_4$ by such steps is \emph{stacked}.

\subsection{A bin represented by a triangular face}
Take a unary instance of \textsc{3-Partition} with $m\ge1$, capacity $B$, and $3m$ positive item sizes $w_j$ satisfying
\[
  B/4<w_j<B/2,\qquad \sum_j w_j=mB.
\]
The problem remains NP-complete with these bounds and unary encoding~\cite{gj}. We first construct a degree list $q$ together with a set $F$ of prescribed edges on its positions.

\begin{figure}[htbp]\centering
\begin{tikzpicture}[x=.82cm,y=.8cm,
  bin vertex/.style={circle,fill=black,inner sep=1.65pt},
  bin cap/.style={circle,draw=black,fill=white,inner sep=2pt},
  prescribed/.style={draw=binPath,line width=1.25pt},
  completion/.style={draw=black!62,line width=.45pt}]
\definecolor{binPath}{HTML}{A4382B}
\coordinate (r) at (0,1.85);
\coordinate (a) at (0,-1.85);
\coordinate (c) at (10,0);
\foreach \i in {1,...,10}
  \coordinate (p\i) at ({10-.91*\i},0);

\foreach \i in {1,...,9}\draw[completion] (r)--(p\i)--(a);
\draw[completion] (a)--(p10) (c)--(p1) (p9)--(p10);
\draw[completion,dashed] (p3)--(p4) (p6)--(p7);
\draw[prescribed] (r)--(a)--(c)--cycle (r)--(p10);
\draw[prescribed] (p1)--(p2)--(p3) (p4)--(p5)--(p6) (p7)--(p8)--(p9);

\foreach \i in {1,...,9}\node[bin vertex] at (p\i) {};
\node[bin cap,label=left:{$y_i$}] at (p10){};
\node[bin vertex,label=above:{$R_i$}] at (r){};
\node[bin vertex,label=below:{$A_i$}] at (a){};
\node[bin vertex,label=right:{$C_i$}] at (c){};
\end{tikzpicture}
\caption{A bin of capacity nine. The nine filled interior vertices form three prescribed paths of three vertices each. Dashed edges join consecutive items. The remaining prescribed edges in the face are its bold boundary and the bold edge from $R_i$ to the open cap $y_i$.}\label{fig:bin}
\end{figure}
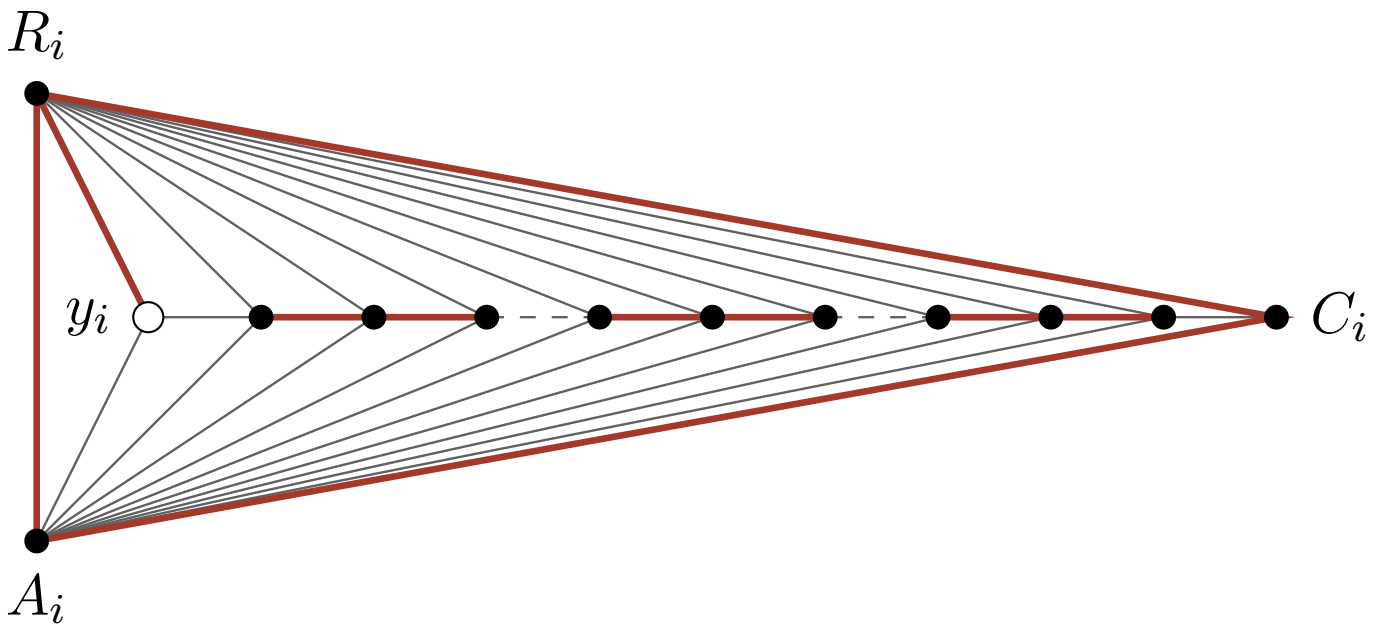

Figure~\ref{fig:bin} shows how one triangular face can accommodate a bin. Beginning in the face $R_iA_iC_i$, insert the $B$ item vertices in succession, each adjacent to $R_i$, $A_i$ and its predecessor, with $C_i$ as the predecessor of the first. A final vertex $y_i$ caps the chain. Each item vertex then has degree four, the cap has degree three, and the three corners gain $B+1,B+1,1$ neighbours, respectively.

For the skeleton $Q$ take the third power of the path on $0,\ldots,4m+2$, a stacked triangulation also used by Maffucci~\cite[Appendix~A]{maffucci}. The first four vertices form a tetrahedron, and each subsequent vertex can be stacked on the face formed by its three immediate predecessors. For bin $i$, where $0\le i<m$, take
\[
 (R_i,A_i,C_i)=(4i,4i+1,4i+3).
\]
This triangle is a face of the initial tetrahedron when $i=0$ and is created at the insertion of $4i+3$ otherwise. Because its three labels are not consecutive, it remains a face throughout the construction. The bin faces use disjoint sets of vertices, and the roots $R_i$ are pairwise nonadjacent in $Q$.

To form $q$, take the degrees in $Q$ with the increments just described at the bin corners. For each item $j$ add a path on $w_j$ new vertices of target degree four, and for each bin add a cap of target degree three. Prescribe the edges of $Q$, the edges within each item path and the edge $R_i y_i$ at each cap. Since $Q$ has maximum degree at most six, the prescribed graph has maximum degree at most seven. The increments at the old vertices total $m(2B+3)$, and the new vertices contribute $4mB+3m$. Thus $q$ has order $s=4m+3+m(B+1)$ and $\vol(q)=6s-12$.

\begin{lemma}\label{lem:bins}
The list $q$ has a planar realization containing $F$ if and only if the items admit a $3$-partition. In that case there is a stacked realization containing $F$. For every source instance, $q$ alone has a stacked realization.
\end{lemma}
\begin{proof}
Given a packing, concatenate the item paths assigned to each bin, preserving the order of the vertices along each path. The construction of Figure~\ref{fig:bin} realizes all the required degrees, and the edges within each item path and at each cap are present. If only the degree list matters, we can distribute the $mB$ item vertices arbitrarily into groups of size $B$ and apply the same construction.

Conversely, suppose a planar realization contains $F$. Its skeleton vertices induce exactly $Q$, since $Q$ already attains Euler's edge bound. Each item path, even one consisting of a single vertex, lies in one open face of $Q$ by connectedness. Any edge from the path to a root ends at a corner of that face. Since the three corners form a triangle and the roots are independent in $Q$, at most one of them is a root. The cap $y_i$ likewise lies in a single open face, with $R_i$ on its boundary by the prescribed edge $R_iy_i$. Thus each new vertex meets at most one root. The increments at the roots sum to $m(B+1)$, the number of new vertices, so every new vertex meets exactly one root, the same one for all vertices of an item. Once its cap is counted, each root has $B$ neighbours left for items. Since $B/4<w_j<B/2$, these neighbours belong to exactly three items.
\end{proof}

\subsection{Reading a layer backwards}
The list $q$ alone always has a stacked realization. To force the edges of $F$ by degrees, we use layers of insertions that record which corners belong to skipped faces. Let $\pi$ have $N\ge4$ positions, all of degree at least three, with $\vol(\pi)=6N-12$. Choose a set $U$ of $3t$ distinct positions. One round produces the list $\pi'$ given by
\begin{equation}\label{eq:lift}
  \pi'_v=2\pi_v-\mathbf1_{v\in U}\quad(v\text{ old}),\qquad
  b=2N-4-t\quad\text{new entries of degree three}.
\end{equation}
Since $3t\le N$ and $N\ge4$, we have $b\ge3$, including when $t=0$. The new order is $N'=3N-4-t>4$ and $\vol(\pi')=6N'-12$. When $t=0$, write $\mathcal F(\pi)=(2\pi,3^{2N-4})$ for this full face insertion.

\begin{lemma}\label{lem:lift}
Let $\pi'$, $U$ and $t$ be as in~\eqref{eq:lift}. From a triangulation of $\pi$ in which $U$ is the union of the corners of $t$ vertex-disjoint faces, inserting one vertex in every other face gives a triangulation of $\pi'$.

Conversely, take \emph{any} planar realization of $\pi'$ and assign its vertices to positions of the same degree in any order. In a sphere embedding the cubic vertices are independent. Deleting all of them leaves a triangulation on the old positions with degrees $\pi$. Precisely $t$ of its faces received no insertion. Their vertex sets are disjoint and have union $U$.
\end{lemma}
\begin{proof}
If we leave the specified faces empty and stack once on each remaining face, each vertex $v$ gains a neighbour in every one of its $\pi_v$ incident faces except for the unique skipped face when $v\in U$. For example, filling three faces of $K_4$ changes $(3^4)$ to $(6,5^3,3^3)$.

For the converse, label a triangulation of $\pi'$ by the indicated positions. Since every old degree is at least five, the new vertices are precisely those of degree three. Two cubic vertices cannot be adjacent in a simple sphere triangulation of order greater than four. Indeed, the two faces at their common edge would give them the same two other neighbours, and the four triangles at the pair would form an entire tetrahedral sphere.

We can therefore delete all the cubic vertices, replacing each three-face fan by its boundary triangle. The remaining triangulation has order $N$, and the deleted fans give distinct restored faces. In fact, two fans on opposite sides of one boundary triangle would fill the sphere and leave only three old vertices. A fan opposite an unfilled face with the same boundary would fill a tetrahedral sphere of order four.

Write $a_v$ for a remaining degree and $h_v$ for the number of incident faces without an insertion. Counting incident edges before deletion gives
\[
  2a_v-h_v=2\pi_v-\mathbf1_{v\in U}.
\]
Since $h_v\ge0$ and degrees are integral, $a_v\ge \pi_v$ at every vertex. Both degree sums equal $6N-12$, so equality holds everywhere and $h_v=\mathbf1_{v\in U}$. Precisely $t$ faces were skipped, each selected position is used once, and the other positions are unused.
\end{proof}

\paragraph{Reading a degree through several layers.}
Fix the total number $R$ of layers. Suppose a position $x$ first appears in round $j$ with degree $a$, where $j=0$ for an original position and $a=3$ for an inserted one. If $D_r$ denotes its degree after round $r$ and $\varepsilon_r=\mathbf1_{x\in U_r}$ for $r>j$, then~\eqref{eq:lift} gives
\[
       D_j=a,\qquad D_r=2D_{r-1}-\varepsilon_r,\qquad
       D_R=2^{R-j}a-\sum_{r=j+1}^{R}2^{R-r}\varepsilon_r.
\]
The last degree determines the entire record. At round $r\ge1$, a value of three identifies a position born in that round, since every surviving old degree is at least five. At any larger value $D$, its parity gives $\varepsilon_r$ and $\lceil D/2\rceil$ gives the previous degree. If the descent reaches round zero, the remaining value is the original degree, which may itself be three. Thus positions of equal final degree have the same birth round, initial degree and subsequent selection bits.

\subsection{Eliminating prescribed edges}
Place a marker over each prescribed edge so that its only two original neighbours are the ends of that edge. To recover their identities from degrees alone, group disjoint edges into matchings and number the edges within each matching. Each application of Lemma~\ref{lem:lift} forces incidences for a selected group of markers. By selecting one binary digit class at a time, we identify both ends of every edge in a logarithmic number of rounds per matching.

\begin{lemma}[Elimination of prescribed edges]\label{lem:eliminate}
Fix a nonnegative integer $c$. Let $q$ be a list on $s\ge4$ positions with $3\le q_v<s$ and $\vol(q)=6s-12$, and let $F$ be a simple graph on the positions with maximum degree at most $c$. In polynomial time one can write an ordinary degree multiset $d$ with all entries at least three and sum $6|d|-12$, where $|d|$ is its number of entries. It has a planar realization if and only if some sphere triangulation with degrees $q_v$ contains all the edges of $F$.

Every such triangulation extends by face insertions to a realization of $d$. From every planar realization of $d$ one can recover, by deleting layers of cubic vertices and identifying the surviving positions, an induced triangulation realizing $(q,F)$. In particular a stacked witness for $(q,F)$ extends to a stacked witness for $d$.
\end{lemma}
\begin{proof}
If $F$ has no edges, use the multiset of entries of $q$. Any planar realization can be given the prescribed positions by assigning its vertices to positions of the same degree. If $|E(F)|>3s-6$ or some $\deg_F(v)>q_v$, we may output $(5^3,3^3)$, whose three universal vertices force $K_{3,3}$. Assume henceforth that neither bound is exceeded and that $F$ has an edge.

\emph{Construction.} Set $\pi^{(0)}=q$. The first three rounds skip no faces. The first divides each original face into three triangles, each incident with exactly one original edge. In the second round a vertex inserted in such a triangle marks the two ends of that edge. The third round supplies a fresh third corner on either side of each edge from a marker to an endpoint. In each later round, the parity of a surviving degree records whether its position was selected, by~\eqref{eq:lift}.

For each edge $e\in E(F)$, reserve a distinct second-round position $m_e$. This is possible for every $s\ge4$, since the second round adds $6s-12$ vertices and $|E(F)|\le 3s-6$. In any triangulation recovered through the three full rounds, the markers are pairwise nonadjacent, and the two original neighbours of each marker are the ends of an original edge. Conversely, if a triangulation with degrees $q$ contains $F$, we may assign $m_e$ to a vertex inserted over $e$ in the second round. Distinct edges use distinct triangles of the first refinement (Figure~\ref{fig:marker}).

\begin{figure}[htbp]\centering
\begin{tikzpicture}[x=1cm,y=1cm]
  \begin{scope}
    \coordinate (l) at (-1.65,0); \coordinate (r) at (1.65,0);
    \coordinate (z) at (0,1.75); \coordinate (s) at (0,.61);
    \draw (l)--(r)--(z)--cycle;
    \draw (l)--(s)--(r) (s)--(z);
    \fill (l) circle (1.7pt) (r) circle (1.7pt) (z) circle (1.7pt);
    \draw[fill=white] (s) circle (2.2pt);
    \node[below] at (l) {$u_i$}; \node[below] at (r) {$v_i$};
    \node[above] at (z) {$z$}; \node[above right,inner sep=2pt] at (s) {$m_i$};
  \end{scope}
  \begin{scope}[xshift=5cm]
    \coordinate (l) at (-1.65,0); \coordinate (r) at (1.65,0);
    \coordinate (s) at (0,1.75); \coordinate (a) at (0,.61);
    \fill[black!14] (l)--(s)--(a)--cycle;
    \draw (l)--(r)--(s)--cycle;
    \draw (l)--(a)--(r) (a)--(s);
    \fill (l) circle (1.7pt) (r) circle (1.7pt);
    \draw[fill=white] (s) circle (2.2pt);
    \draw[fill=white] (a) ++(-.07,-.07) rectangle ++(.14,.14);
    \node[below] at (l) {$u_i$}; \node[below] at (r) {$v_i$};
    \node[above] at (s) {$m_i$}; \node[above right,inner sep=2pt] at (a) {$a$};
  \end{scope}
\end{tikzpicture}
\caption{Round two inserts $m_i$ in the first-round face $u_iv_i z$. The right panel enlarges $u_iv_im_i$ after round three inserts $a$ (square). Skipping the shaded face in the first test marks $u_i,m_i,a$. Later bit tests recover the marker--endpoint pairs.}\label{fig:marker}
\end{figure}
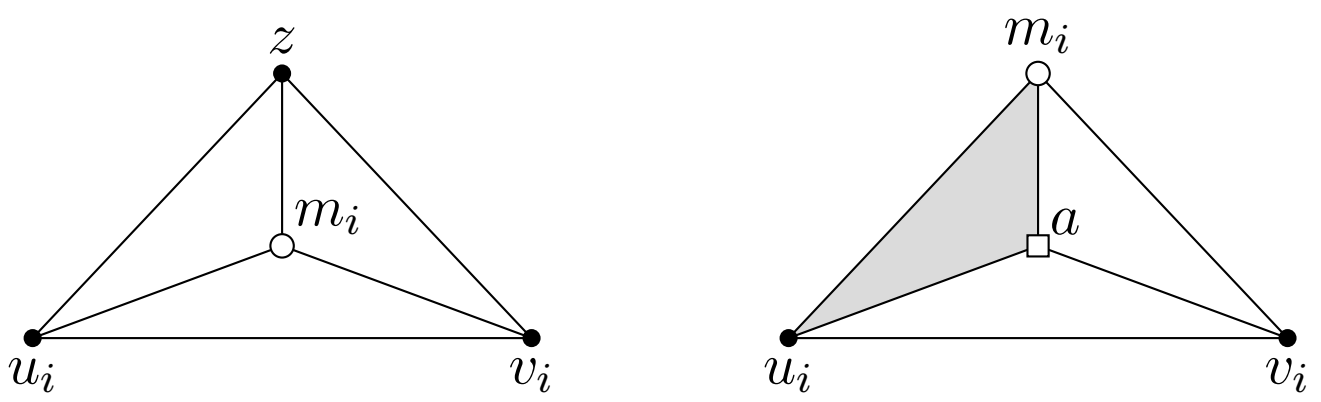

A greedy edge-colouring partitions $F$ into at most $2c-1$ matchings, since at most $2c-2$ other edges meet the endpoints of an edge. Take the edges in lexicographic order and assign the first available colour. In each nonempty matching write the edges in that order as $u_iv_i$ with $u_i<v_i$, $0\le i<k$, and write $m_i$ for their markers.

For this matching first test the whole index set on the $u$-side and on the $v$-side. Next write each index in $\lceil\log_2 k\rceil$ bits. For each bit, test its nonempty zero class and its nonempty one class, each on both sides. For $k=1$ the two whole-set tests suffice. A test of $J$ on the side $\xi\in\{u,v\}$ specifies one further round $r$ with
\[
   U_r=\{m_i:i\in J\}\ \cup\ \{\xi_i:i\in J\}\ \cup\ A_r,
\]
where $A_r$ consists of the first $|J|$ positions born in round $r-1$. Indeed, $|J|\le s/2$, while the preceding round skipped at most $s/2$ faces on at least $s$ vertices and so inserted at least $2s-4-s/2\ge s/2$ vertices. Visit the matchings by colour, the bits from low to high, and the zero class before the one class, taking the $u$-side before the $v$-side in each test pair. In every round compute $\pi^{(r)}$ from $\pi^{(r-1)}$ and $U_r$ by~\eqref{eq:lift}, appending the new positions at the end. The output $d$ is the multiset of entries of $\pi^{(R)}$ after the last round.

\emph{Extending a prescribed realization.}
Take a triangulation of $q$ containing $F$ and choose the second-round marker of each prescribed edge as above. Suppose round $r$ tests $J$ on the $\xi$-side. On each edge $m_i\xi_i$ with $i\in J$, at least one incident face has a third corner born in round $r-1$. In the first test both sides were just subdivided. At each later test the preceding round skipped vertex-disjoint faces, so it could not skip both sides of this edge. The edges $m_i\xi_i$ are pairwise disjoint. A fresh vertex has only three older neighbours and thus cannot be the third corner for two of these edges. Assign the names in $A_r$ to the chosen third corners and skip the corresponding faces. These fresh vertices all have degree three, and this is their first use in a test. After round $r$ they are too old to serve as auxiliaries again. All earlier assignments, in particular those of the markers, remain fixed. Iterating constructs a realization of $d$ by face insertions and preserves stackedness.

\emph{Recovering all prescribed edges.}
Fix a planar realization of $d$ and a degree-preserving bijection $\phi$ from its vertices to the positions of $\pi^{(R)}$. As we apply Lemma~\ref{lem:lift} backwards, we restrict this same bijection to the surviving vertices. Their degree histories determine which positions to delete and which were selected in each round. The original positions thus induce a single triangulation of $q$ to which all tests refer.

In the round testing $J$ on the $\xi$-side, the $|J|$ skipped faces partition $U_r$. Both the markers and the fresh auxiliaries are independent sets of size $|J|$, so each face contains one of each and one selected endpoint. Its marker--endpoint edge persists when we delete the later layers. Applying the two whole-set tests shows that $m_i$ has an original neighbour on each side of its matching. These neighbours are unique because the two sides are disjoint and $m_i$ has only two original neighbours. For each bit of $i$, the test of its digit class forces the neighbour on the $\xi$-side to have that same bit. Thus the neighbours of $m_i$ are $u_i$ and $v_i$, and the original edge over which the marker was inserted is $u_iv_i$. This holds for every prescribed edge in the triangulation recovered using $\phi$.

\emph{Output size.}
A matching of size $k$ uses at most $2+4\lceil\log_2 k\rceil$ tests. In the nonempty case $c\ge1$, so
\[
  R\le 3+(2c-1)(2+4\lceil\log_2 s\rceil),\qquad |d|\le 3^R s.
\]
Each layer at most doubles the largest degree. The entries thus have at most $O(R+\log s)$ bits, and generating the successive lists takes polynomial time for fixed $c$. In the bin construction $c=7$, giving the upper bound $|d|=O(s^{1+52\log_2 3})$, with exponent about $83.42$. Here $s$ measures the prescribed-edge instance constructed from unary \textsc{3-Partition}.
\end{proof}

\subsection{Completing the reduction}
We apply Lemma~\ref{lem:eliminate} to the bin construction with $c=7$. To check the degree bound, recall that the bin faces are vertex-disjoint, so a skeleton vertex of degree at most six gains at most $B+1$ neighbours. Therefore $q_v\le B+7<s=mB+5m+3$ for every skeleton vertex, including when $m=1$, while the remaining entries are three or four. All degrees are at least three. Since $s=O(mB)$ is polynomial in the unary input length, Lemmas~\ref{lem:bins} and~\ref{lem:eliminate} give a polynomial reduction from \textsc{3-Partition} with a stacked realization for every yes-instance. Invalid inputs may be mapped to $(5^3,3^3)$, realized by the $3$-connected join of a triangle and three isolated vertices.

The output has $n$ entries of sum $6n-12$, so its unary representation, including separators, uses only $O(n)$ symbols. This proves strong NP-hardness. For membership in NP, a labelled edge list allows us to check simplicity and the input degrees directly and test planarity in time linear in the size of the graph~\cite{ht}.

To prove the assertion about no-instances, start with the unrestricted stacked realization of $q$, which is $3$-connected because each insertion from $K_4$ joins a new vertex to three old ones. More generally, given a $3$-connected simple realization of a current list $\pi$, we shall extend it by attaching $\pi_v-\mathbf1_{v\in U}$ new neighbours to each old vertex $v$. The demands sum to $3b$, and no demand exceeds $b$, since $3t\le N$ and $N\ge5$ give $b\ge N-1\ge \pi_v$. Repeat the $b$ new names three times in the same cyclic order and cut the resulting word into blocks of the required lengths, one per old vertex. No block repeats a name, so every new vertex has three distinct old neighbours. After deleting at most two vertices the surviving old graph is still connected and contains a neighbour of every surviving new vertex, as required for Theorem~\ref{th:hard}.

The stacked witnesses give NP-completeness for planar $3$-tree degree lists (Corollary~\ref{cor:stacked}). For explicit lists with a fixed number of distinct degrees, the face recurrence in Appendix~\ref{app:stacked} recognizes stacked realizations in polynomial time.

\section{The asymptotic subdivision budget}\label{sec:open-budget}
Consider the graphic lists $h$ of length $p\ge8$, minimum degree at least three and sum $6p-12$. On this class even testing $T(h)\le p/29-1$ is strongly NP-complete when $29\mid p$ (Theorem~\ref{th:compact-fixed-twos}). To measure how many twos suffice for every such list of order $p$, put $M(p)=\max_h T(h)$. Proposition~\ref{prop:compact-spectrum} and Corollary~\ref{sp:cor:eight} give
\[
 \left\lfloor\frac{p-4}{2}\right\rfloor \ \le\ M(p)\ \le\ p-6 .
\]
The symmetric lists $((p-1)^3,3^{p-3})$ attain the lower bound (Appendix~\ref{app:spectrum}).

\paragraph{Open problem.}
Determine the leading asymptotic term of $M(p)$. Are the three-pole lists asymptotically extremal, in the sense that $M(p)=p/2+o(p)$?

\clearpage
\appendix
\section{Stacked realizations and subdivision budgets}\label{app:consequences}
\subsection{Stacked triangulations}\label{app:stacked}
\begin{corollary}\label{cor:stacked}
Recognition of explicitly listed degree sequences is NP-complete both for maximal planar graphs and for planar $3$-trees.
\end{corollary}
\begin{proof}
At orders $n\ge4$, planar $3$-trees are precisely stacked triangulations~\cite[Sec.~2]{kv}.
The reduction gives stacked realizations on positive instances and excludes every planar realization on negative ones.
For NP membership, verify the degrees and either a planar edge list with $3n-6$ edges or a stacking order from $K_4$.
Smaller orders are immediate.
\end{proof}

For an explicit list of length $n$ with $j$ distinct degrees, stacked recognition takes
$(n+1)^{3j}n^{O(1)}$ time after input. For each submultiset of interior degrees,
tabulate the possible triples of new-neighbour counts at the ordered corners of a triangular face.
The empty filling gives $(0,0,0)$. A first insertion of degree $k$ splits the face into
three. Their counts at the new vertex sum to $k-3$ and each old corner gains one plus its
two incident counts. Conversely, glue the three fillings in their respective faces.
At $K_4$, choose four corner degrees and combine four fillings whose counts at each
corner sum to its degree minus three. Among at most $(n+1)^j$ submultisets,
three child multisets and $k$ determine their parent, and at $K_4$ three multisets
and the corner degrees determine the fourth. Each count is at most $n$.

\subsection{Hardness with a prescribed subdivision budget}
\begin{theorem}\label{th:compact-fixed-twos}
For every fixed integer $r\ge1$, deciding whether $T(h)\le r$ is strongly
NP-complete even for graphic lists $h$ of length $p$, minimum degree at
least three and sum $6p-12$. On the same class with $29\mid p$, deciding
whether $T(h)\le p/29-1$ is also strongly NP-complete.
\end{theorem}
\begin{proof}
Reduce from the graphic maximal-sum instances $d$ of Theorem~\ref{th:hard}
with $n=|d|\ge r+3$. Omitting finitely many orders preserves hardness.
Let $\pi^{(1)}=\mathcal F(d)$ and $\pi^{(2)}=\mathcal F(\pi^{(1)})$
be two rounds of full face insertion, and put $N=|\pi^{(2)}|$.
Mark $r$ original positions and $r$ positions born in the first round,
with union $U$. Set $b=2N-4+2r$ and take
\[
 h=\bigl((2\pi^{(2)}_i+3\mathbf1_{i\in U})_{i=1}^N,\,3^b\bigr).
\]
This list has $p=N+b$ entries of sum $6p-12$. At each marked position $i$,
the edge count alone will allow the recovered degree to exceed
$\pi^{(2)}_i$ by one. The full layer beneath it forces both degrees to be
even, ruling out that extra unit.

In a triangulation of $d$, Hall's theorem matches the marked original
vertices to distinct incident faces. Each vertex meets at least
three faces and each face meets at most three marked vertices.
Assign the marked new positions to the corresponding face centers in the first
round. Two further full insertions preserve the resulting matching and
give the list $\mathcal F(\pi^{(2)})$. Along each matched edge $xy$ draw
three two-edge paths and join
two consecutive internal vertices. This raises both endpoint degrees by
three and adds two cubic vertices and one degree-two vertex per edge,
realizing $(h,2^r)$.

Conversely, suppose $(h,2^s)$ has a planar realization for some $s\ge0$.
The $N$ vertices from $\pi^{(2)}$ induce at most $3N-6$ edges, so at least
$2\sum_i\pi^{(2)}_i+6r-2(3N-6)=3b$ edges join them to the remaining vertices.
These crossing edges form a simple bipartite planar graph, giving
$3b\le2(N+b+s)-4$. Since $3b=2(N+b+r)-4$, we have $s\ge r$.
It remains to consider $s=r$. Both edge bounds are then tight, so the old vertices induce a triangulation.
The crossing graph has $2(N+b+r)-4$ edges, the largest possible number
for any bipartite planar graph on these vertices. Move a single old vertex
$i$ to the other side of the cut. If its degree in the old triangulation
is $g_i$, the move adds $g_i$ crossing edges and removes
$\deg(i)-g_i$. The new cut cannot be larger, so
$2g_i\le\deg(i)=2\pi^{(2)}_i+3\mathbf1_{i\in U}$.

The positions born in round two are exactly the old positions of target
degree six in $h$. They were unmarked and cubic in $\pi^{(2)}$, whereas
the positions inherited from $\pi^{(1)}$ had degree at least six there.
The inequality gives $g_i=3$ at the round-two positions, since a triangulation
has minimum degree three. Deleting these $2|\pi^{(1)}|-4$ cubic vertices
restores distinct faces as in Lemma~\ref{lem:lift}, regardless of any other
cubic vertices. Their number is the full face count of the remaining
triangulation, so each surviving vertex had one deleted neighbour in every
incident face. At each such vertex both $g_i$ and its round-two target
$\pi^{(2)}_i$ are even. At a marked position the inequality
gives $g_i\le\pi^{(2)}_i+1$, which by parity improves to
$g_i\le\pi^{(2)}_i$. The same bound holds at the unmarked positions.
The two degree sums equal $6N-12$, so $g=\pi^{(2)}$. Reversing the two full
insertions now gives a triangulation of $d$.

Finally $h$ is graphic for every source instance. The cyclic assignment
in the proof of Theorem~\ref{th:hard} realizes each $\mathcal F(x)$ from a simple
realization of $x$ while preserving the old induced graph. After the first round each marked
original vertex has at least $(2n-4)-(n-1)\ge r$ new nonneighbours.
Choose distinct ones greedily and assign them the marked new positions.
After two further full rounds, add each such missing edge
$xy$ and two adjacent vertices both joined to $x,y$. This realizes $h$. Since $|\mathcal F(x)|=3|x|-4$, the output
core has order $p=27n-52+2r$. Choosing $r=n-3$ gives $p=29(r+1)$.
Both choices of $r$ yield outputs of linear order and linear unary length.
A planar realization of $(h,2^r)$ certifies membership in NP.
\end{proof}

\subsection{The range of subdivision budgets at fixed order}\label{app:spectrum}
\begin{proof}[Proof of Proposition~\ref{prop:compact-spectrum}]
The vertices $A,B,C$ of the first three entries send at least
$2(p-1)+(q-1)-6=2p+q-9$ edges across their cut in any planar realization
of $(h,2^t)$. The bipartite planar bound $2(p+t)-4$ gives $t\ge\lceil(q-5)/2\rceil=m$.

Start with $K_4$ on $A,B,C,u$. Figure~\ref{fig:compact-six} (right) shows the construction for $m=1$ and $k=0$. Inside the face $ABC$ draw $m$ internally
disjoint two-edge paths along each side. Order the internal
vertices of the $AB$ and $BC$ strips from $AB$ inwards and then out to $BC$.
Join them in consecutive pairs, each in its own quadrilateral or in the central
face. The $2m$ paired vertices are cubic, the $m$ along $CA$ have degree two,
and each corner has degree $3+2m=q-1$.

Stack $k$ vertices starting in $ABu$, using the new face at $AB$ each time.
Each insertion raises both $A,B$ by one, raises the preceding cubic vertex
to degree four, and adds a cubic vertex. This gives $(h,2^m)$.
To realize $h$ alone, join a triangle to $q-3$ independent vertices and make
the same additions starting at any independent vertex. Its degrees sum to $6p-12$.
\end{proof}

The symmetric list $g_p=((p-1)^3,3^{p-3})$ satisfies
$T(g_p)=\lceil(p-5)/2\rceil$ for every $p\ge8$.
Even orders follow by taking $k=0$. For odd $p=2m+5$ the same cut gives
$t\ge m$. Before pairing the strip vertices, add a vertex adjacent to $A,B,C$
in the central face. Within the strips there are
$3\lfloor m/2\rfloor\ge m$ disjoint consecutive pairs since $m\ge2$.
Joining any $m$ pairs in their quadrilaterals leaves $m$ twos and gives
each corner degree $4+2m=p-1$.

\section{Derivation of the arithmetic criterion}\label{app:arithmetic-proof}
\begin{proof}[Proof of Theorem~\ref{th:compact-arithmetic}]
Take a Schur complement of $I+C$ and apply the weighted matrix-tree theorem~\cite{klee} with edge weights $c_e/(1+c_e)$. Writing $T$ for a spanning tree of $K_n$, the determinant expands as
\[
 \sum_T\prod_{e\in T}c_e\prod_{e\notin T}(1+c_e).
\]
 Each simple graph $G$ thus has weight equal to its number of spanning trees. Differentiation at zero keeps precisely the graphs with degree list $d$.

In the trace expansion~\cite{kang}, matching derivative positions to incident edge ends chooses a cyclic order at each vertex and a starting end. Write $\sigma$ for the permutation of ends in these orders and $\alpha$ for edge reversal. Since $\operatorname{tr}(X_iX_j)$ pairs matrix entries with reversed indices, the free indices run once per cycle of $\sigma\alpha$, the face permutation. A connected rotation system with $b$ faces contributes a positive multiple of $q^b$. Euler's formula gives $1\le b=f-2g$ for genus $g$. Hence the constant term vanishes, and the $f$th forward difference at zero is $f!$ times the coefficient of $q^f$, which counts the weighted spherical systems.

The determinant counts only connected graphs. If a planar realization is disconnected, $m\ge n$ guarantees a cycle in one component. Expose a cycle edge and an edge of another component on their outer faces. Delete both and reconnect their endpoints across the components without crossings. This preserves degrees and merges the components even if the second edge was a bridge. Iteration gives the connected spherical realization counted above.
\end{proof}

\section{Proof of the sparse boundary theorems}\label{sp:app:proofs}
In the notation of Section~\ref{sp:sec:results} put
$\delta=4n-2\ell-S$ and $j=\delta/2$. The two boundary layers have $j=1,0$, respectively.
We first treat leaf-free lists $(h,2^t)$, writing
$a,b,c,\ldots$ for the core degrees in descending order, $p$ for their
number and $H$ for their sum. The budget and the degree bound are
\begin{equation}\label{sp:eq:budget}
 t=H/2-2p+j,\qquad U_{\max}=p+t-1.
\end{equation}
By Lemma~\ref{lem:compact-support}, $(h,2^t)$ is planaric exactly when
some plane multigraph $M$ of degrees $h$ satisfies $s(M)-\lambda(M)\ge2p-j$.
Here $s(M)$ counts the distinct nonloop adjacencies and $\lambda(M)$ counts loops with multiplicity.

Section~\ref{sp:sec:tools} gives the local constructions for extending a
smaller realization. They yield all leaf-free exceptions with $p\le7$ in
Section~\ref{sp:sec:small} and the base cases for induction. For $p\ge8$,
Section~\ref{sp:sec:large} reduces repeated core degrees in pairs and restores
degrees four and three individually. Section~\ref{sp:sec:leaves} then restores
the leaves and tracks which exceptions survive.

We shall also use a cut count with leaves present. On $j=0$ let $C$
be the core and $R$ its complement. Subtracting the two parts' degree
sums gives $e(C)-e(R)=2p$. Hence $p\ge5$ for any nonempty realization,
and $p=5$ forces $K_5$. For $p=6$, $2e(C)\ge24$.
Four or more threes would give $2e(C)\le4\cdot3+2\cdot5=22$.
Exactly three threes force equality in $2e(C)\le3\cdot3+3\cdot5=24$:
the induced graph on $C$ has three vertices adjacent to every other
core vertex, so it contains $K_{3,3}$.

\subsection{Construction rules}\label{sp:sec:tools}
\begin{lemma}\label{sp:lem:eg}
Suppose $p\ge5$, $H$ is even, $j$ is a nonnegative integer, $p+j\ge6$,
and $t=H/2-2p+j\ge0$. Then $(h,2^t)$ is graphic exactly when $a\le U_{\max}$.
In that case $H-a-b\ge2p+2-2j$.
\end{lemma}
\begin{proof}
In the Erd\H{o}s--Gallai test~\cite{sp:eg} the first inequality is $a\le U_{\max}$
and the second follows from $H-a-b\ge H-2U_{\max}=2p+2-2j$.
For $3\le k\le p$ the right side minus the left is at least
$k(k-1)+6(p-k)+2t-H=k^2-7k+2p+2j\ge0$.
For $k\ge p$ and $k\ge3$ the difference equals $k(k-5)+2j\ge0$.
\end{proof}

\begin{lemma}\label{sp:lem:insert}
In a simple plane realization of minimum degree at least two, a prescribed vertex
can gain one neighbor while a three is added to the list. A four can
be added with every old degree fixed if $m<3n-6$.
\end{lemma}
\begin{proof}
For the first move take an edge $wu$ at the prescribed vertex $w$.
Subdivide the next edge at $u$ and join its new vertex to $w$ through
the intervening sector. For the second move take two disjoint edges
on one face, join their four ends to a new vertex in that face and erase
both edges. Edges in different components can be placed on the outer face.
In a connected realization Euler gives a nontriangular face. Its boundary
walk has three distinct vertices in every consecutive triple (minimum
degree two). If every consecutive quadruple repeats a vertex, the walk
is periodic of period three and uses a directed edge twice unless it
is a triangle. Hence the required pair exists.
\end{proof}

\begin{lemma}\label{sp:lem:pair}
From a planar realization of $(h,2^s)$ with $s\ge2$, one can realize
$(h,g,g,2^{s+g-4})$ for $g\ge3$.
\end{lemma}
\begin{proof}
If a two has nonadjacent neighbors, suppress it. At a remaining
two $z$ subdivide an incident edge by $x$ and add $g-2$ parallel
two-edge paths alongside $zx$. Otherwise delete a two whose neighbors
are $u,v$, glue a copy of $K_4$ along $uv$ in a thin sector, and erase
$uv$. Along the edge joining its two fresh vertices add $g-3$ parallel
two-edge paths. Both operations preserve the other prescribed degrees.
\end{proof}

\begin{lemma}\label{sp:lem:cone}
If $p\ge6$, every $h_i\ge5$ and $t+5-j\le a\le U_{\max}$, then $(h,2^t)$
is planaric on either boundary.
\end{lemma}
\begin{proof}
The square of a path on $q=p-1$ vertices is outerplanar with $2q-3$
edges and maximum degree four. Cone it and repeat the spokes to fill
all rim targets. The apex has degree $H-a-4q+6$, so $a-t-5+j$ loops
at it complete the degrees. The support has $3p-6$ edges and the
cost is $a-p+1\le t$.
\end{proof}

We use two edge-weighting rules. On a complete multipartite graph,
nonnegative increments summing to $2s$ are attainable by copying edges
exactly when each part needs at most $s$: pair a unit in each of two
parts of largest positive demand. All parts of demand $s$ are chosen,
so induction applies. On the planar double fan $K_2\vee P_q$, let
$A\ge B$ be the two poles' residual demands and $R$ the total residual
path demand. Assume even total, all demands nonnegative, and $R\ge A-B$.
If one can place $\max\{0,(R-A-B)/2\}$ path-edge copies within the
path capacities, the degrees can be completed. For $R\ge A+B$ place
exactly that many copies; for $R<A+B$ copy the pole edge $(A+B-R)/2\le B$
times. The remaining path demand equals the total remaining pole demand
and all spokes are available.

\begin{lemma}\label{sp:lem:sixfour}
Let $T(h)$ denote the minimum number of additional twos. For $h=(a,b,c,d,e,f)$ with $f\ge4$ and $H$ even,
$T(h)=\max\{H/2-12,a-5\}$, except that $T(6,4^5)=2$.
\end{lemma}
\begin{proof}
The two lower bounds follow from the maximum number of edges on a
six-vertex planar support and from the maximum degree of a simple graph.
If $a+f\le H/2-4$, take $K_{2,2,2}$ with parts $(a,f),(b,e),(c,d)$.
Each vertex needs its degree minus four in extra copies. For the last
two parts, $H-2(b+e)\ge d+f\ge8$ and $H-2(c+d)\ge e+f\ge8$.
Thus every part needs at most $(H-24)/2$, and the multipartite rule
gives a completion at cost $H/2-12$.

Suppose $b\ge5$ and $a\ge H/2-7$. To give the cone apex degree $a$,
we need a base on $a$ vertices. Start with a triangulated pentagon of
degrees $4,3,3,2,2$, placing $b$ at its vertex of degree four.
Attach $H-a-19$ leaves to bring the five degrees to $b-1,\ldots,f-1$.
The base still needs $(a-5)-(H-a-19)=2a-H+14$ vertices. This number
is nonnegative by the bound on $a$ and even since $H$ is even.
Add them in disjoint pairs joined by an edge. Coning this outerplane
base turns all $a-5$ vertices outside the pentagon into degree-two vertices.
If $b=4$, the five smallest targets are all four, so $H=a+20$ forces
$a$ even. We treat $a=6$ at the end of the proof. For $a=4$ use the
octahedron. For $a\ge8$,
add one chord to a pentagon and attach a leaf to each of its three
vertices of degree two. Add isolated edges on $a-8$ vertices and cone.

In the remaining case with $b\ge5$, we have $a+f\ge H/2-3$ and
$a\le H/2-8$. Put $a,b$ at the poles of $K_2\vee P_4$.
Here $A=a-5$, $B=b-5$, $R=c+d+e+f-14$, and $R-(A-B)=H-2a-14\ge2$.
If $a+b\ge H/2-2$, then $R\le A+B$ and the double-fan rule applies.
Otherwise $a+b=H/2-3=a+f$, whence $b=c=d=e=f\ge5$.
Now $R=A+B+2$ and any path edge accommodates the one extra copy needed.

For $(6,4^5,2)$ the six-valent vertex is universal. Its deletion in
a planar drawing would leave all the other vertices on one face and
give the outerplanar sequence $(3^5,1)$. Deleting its leaf gives a
five-vertex, seven-edge outerplanar graph of maximum degree at most
three. Such a graph triangulates a pentagon, whose two diagonals
share a vertex of degree four. Conversely, attaching a triangle at
one vertex of the octahedron realizes $(6,4^5,2^2)$.
\end{proof}

\subsection{At most seven vertices of degree at least three}\label{sp:sec:small}

\begin{proposition}\label{sp:prop:small}
For a leaf-free graphic sequence on either boundary with $p\le7$,
the nonplanaric sequences are as follows. For $\delta=2$ the only one is
$(4^5,2)$. For $\delta=0$, every sequence with $p=5$ is nonplanaric.
With $p=6$ the exceptions are the lists with exactly three core entries equal to three,
together with $(5,4^4,3)$, $(5^5,3,2^2)$ and $(6,4^5,2)$.
With $p=7$ they are $(4^7)$, $(6,5^2,3^4)$ and $(6^3,3^4,2)$.
\end{proposition}

The support inequality $s(M)-\lambda(M)\ge2p-j$ and Euler's bound
$s(M)\le3p-6$ leave at most $p-6+j$ for the sum of the missing support
edges and the loops when $p\ge3$.
\paragraph{Five core vertices}

For $p\le4$, or $p=5$ with some core degree three, no realization
can contain a subdivision of $K_5$ or $K_{3,3}$, by counting branch
vertices of the required degrees. On the upper boundary,
the core cut count rules out $p\le4$ and forces $K_5$ on the core when $p=5$.

On the lower boundary with $p=5$, assume all five degrees are at least four.
Delete from $K_5$ the edge between the vertices of targets $d,e$.
The resulting graph is complete multipartite, with these two vertices
in one part and the vertices of targets $a,b,c$ in three singleton parts.
The increments needed are $a-4,b-4,c-4,d-3,e-3$ and their sum is
$2t=H-18$. Each singleton part needs at most $a-4\le t$.
The last part needs $d+e-6\le t$ unless all entries are four:
indeed $H-2(d+e)=(a-d)+(b-e)+c\ge4$, with equality only in that
case, and its value is even. The multipartite rule supplies the
extra copies on this nine-edge support. When $h=(4^5)$, $t=1$ and the complement has degree
sequence $(3,1^5)$, realized only by $K_{1,3}\sqcup K_2$. The original
graph is $K_5$ with one edge subdivided.

\paragraph{Six core vertices}

Lemma~\ref{sp:lem:sixfour} treats cores of minimum degree at least four
on both boundaries. Write $r$ for the number of core entries equal to
three. On the upper boundary $t=H/2-12$. A planar realization gives
$s(M)-\lambda(M)\ge12$. A simple planar support on six vertices has at most
twelve edges, so $s(M)=12$ and $\lambda(M)=0$. Such a triangulation is
either the octahedron or $K_2\vee P_4$. Indeed, minimum degree four forces
regularity and a perfect matching in the complement. Otherwise remove a
degree-three vertex. The resulting triangulation is $K_5$ minus an edge,
the triangular bipyramid, all of whose faces are equivalent. Inserting
a vertex in one face gives $K_2\vee P_4$. The core cut count
settles $r\ge3$, so it remains to consider $r=2$ and $r=1$.

For $r=2$ let $h=(a,b,c,d,3,3)$. We have $b\ge5$, since $b=4$ would
force $a\le4$ by the degree bound, and then $H<24$. Place both
threes at the ends of $P_4$. The largest number of additional path
edges is $d-4$. In the double-fan rule, $A=a-5$, $B=b-5$,
$R\ge A-B$ is the degree bound $a\le H/2-7$, and
$(R-A-B)/2=(c+d-a-b+2)/2$. This is at most $d-4$ exactly when
$(a-c)+b+d\ge10$. Since $a\ge c$, $b\ge5$ and $d\ge4$, the left
side is at least nine, and it has the parity of $H$.

For $r=1$ write $h=(a,b,c,d,e,3)$. When $b=4$, the degree bound and
$H\ge24$ force $h=(5,4^4,3)$. The octahedron has minimum degree four,
and the double fan requires two targets of degree at least five.
For $b\ge5$ order the path capacities as $(0,d-4,c-4,e-3)$.
Only the two edges incident with the third path vertex can take extra
copies. They share its $c-4$ units of capacity, and their other ends
supply $d+e-7$ units between them. The maximum number of copies is
therefore $\min\{c-4,d+e-7\}$.
We have $R-(A-B)=H-2a-14\ge0$, as required by the double-fan rule.
The quantity $(R-A-B)/2=(c+d+e-a-b-1)/2$ is at most
$d+e-7$, since $(a-c)+b+d+e\ge 0+5+4+4=13$. It is at most $c-4$ provided
$a+(b-d)+(c-e)\ge7$. This quantity is odd and at least five. Equality
with five forces $h=(5^5,3)$. Both poles are then saturated. The cubic
target must occupy a path endpoint, so up to reversal the capacities are
$(0,1,1,2)$. They require two extra copies but permit only one.

On the lower boundary $t=H/2-11$. Suppose $1\le r\le3$. If $a=4$,
then $H=24-r$ is even, so $r=2$ and deleting an octahedron edge
realizes $(4^4,3^2)$. For $a\ge5$ all three cases use the cone with
apex $x_0$ over the cycle $x_1ux_2vwx_1$ with chord $x_1x_2$. Its degrees in order $x_0,x_1,x_2,u,v,w$
are $5,4,4,3,3,3$. Give $x_0,x_1,x_2$ the targets $a,b,c$.
For $r=2$ put $d$ at $u$, and for $r=1$ put $e,d$ at $v,w$.
All remaining targets are three. Choose the extra edge copies from the
complete multipartite subgraph indicated for each $r$.

\[
\begin{array}{c@{\quad}l@{\qquad}l}
r & \mbox{Parts} & \mbox{Remaining demand in each part}\\
\hline
3 & \{x_0\},\{x_1\},\{x_2\} & a-5,\ b-4,\ c-4\\
2 & \{x_0\},\{x_1\},\{x_2\},\{u\} & a-5,\ b-4,\ c-4,\ d-3\\
1 & \{x_0\},\{x_1,v\},\{x_2,w\} & a-5,\ b+e-7,\ c+d-7
\end{array}
\]

The demands sum to $H-22=2t$, so the multipartite rule requires
only that each part have demand at most $t$.
The degree bound gives $a-5\le t$ throughout. For $r=2,3$ we have
$H-2b\ge c+9\ge13$. As $H$ is even, $H-2b\ge14$, or $b-4\le t$,
which also bounds $c-4$. For $r=2$ we have $H-2d\ge a+b+6\ge15$,
so $H-2d\ge16$, or $d-3\le t$. For $r=1$ we have
$H-2(b+e)=(a-b)+(d-e)+c+3\ge7$ and
$H-2(c+d)=(a-c)+(b-d)+e+3\ge7$. Both are even, so both are at least
eight, as required. Pairing the demands gives $t$ extra edge copies,
each of which we subdivide once.

For $r=4$ the degree bound forces $h=(a,a,3^4)$
with $t=a-5$. The graph $K_2\vee(2K_2\sqcup tK_1)$ realizes it and is
a subgraph of a double fan. For $r\ge5$, the sum and maximum degree
requirements are incompatible: for $(a,3^5)$ they give $a\ge7$ and
$a\le3$, respectively.

\paragraph{Seven core vertices on the upper boundary}

First suppose the minimum target $g$ is at least five, so $t=H/2-14$.
If $a=U_{\max}$, start with a hexagon carrying two noncrossing chords with
disjoint endpoints. Attach leaves to its vertices until each has its
target degree minus one. The number attached is $H-a-22=a-6=t$.
Coning gives the required realization.

Otherwise $a\le H/2-9$. If $a-g\ge2$, remove $g$ and replace
$a$ by $a-g+4$ in the core. The six remaining targets have sum
$K=H-2g+4$ and maximum at most $K/2-7$. For the changed target this
follows from $a\le H/2-9$. For the unchanged targets it suffices to bound $b$:
we need $H-2(b+g)\ge10$, and $H-2(b+g)=(a-b)+(c-g)+d+e+f\ge3g\ge15$.
All six targets are at least five, so Lemma~\ref{sp:lem:sixfour} gives
a simple plane realization with $K/2-12$ twos. Suppress its degree-two
paths to obtain a plane multigraph $M$ with $s(M)-\lambda(M)\ge12$.
Its simple support has at most twelve edges, so it is a triangulation
and $M$ has no loops.
The vertex $w$ of target $a-g+4\ge6$ has at most five neighbors,
so some edge $wu$ is repeated. Fix a plane embedding of the support and let $v$
immediately follow $u$ in the cyclic order at $w$.
Draw the parallel copies in narrow bundles and take the
two facing outermost copies of $wu$ and $wv$. In the
sector between the two edges, replace both copies by $xu,xv$ and
$g-2$ copies of $xw$ for a new vertex $x$. The degree of $w$ rises
by $g-4$ and the new vertex has degree $g$. Another copy of $wu$
survives, so only $wv$ can disappear from the support. The three new
edges $xu,xv,xw$ bring the support to at least fourteen edges, and its
subdivision cost is at most $H/2-14=t$.

For $a-g\le1$ use the pentagonal bipyramid with poles $0,1$ and equator $23456$ in cyclic order. Doubling the Hamilton cycle $02314560$ raises every degree by two, so minima five and six suffice. Even $H$ requires respectively an odd number of sixes or an even number of sevens. Each $ij$ below adds a copy of edge $ij$. The fifteen support edges leave $H/2-15=t-1$ extra copies. Subdivide each once, then subdivide one more edge.

\[
\begin{array}{c|l}
\text{Targets in vertex order}&\text{Extra edge copies}\\
\hline
(6,5^6)&06,23,45\\
(6^3,5^4)&06,12,23,45\\
(6^5,5^2)&06,14,45,23,23\\
(6^7)&06,16,23,23,45,45\\
(7^2,6^5)&06,06,23,23,14,15,45\\
(7^4,6^3)&06,06,23,23,23,14,15,45\\
(7^6,6)&06,06,45,45,23,23,23,14,15
\end{array}
\]

If the minimum is four, the construction for $a=U_{\max}$ still works. For
$a<U_{\max}$ remove one four. The residual remains graphic by
Lemma~\ref{sp:lem:eg}. On the upper boundary the residual has $m'=2n'$,
so $m'<3n'-6$ exactly when $n'\ge7$. Lemma~\ref{sp:lem:insert}
then restores the four in any planar residual. If $n'=6$, all six vertices
are core vertices of degree at least four and their degrees average four.
This gives $(4^6)$ and hence the parent $(4^7)$. Among the six-core
residuals with all core degrees at least four, Lemma~\ref{sp:lem:sixfour}
leaves only $(6,4^5,2)$ as nonplanaric. Table~\ref{sp:tab:rot-orders} realizes
its parent $(6,4^6,2)$.

For $(4^7)$ there is a short face count (see also~\cite{sh}).
A realization would be connected since every component has at least five
vertices. Its even degrees give a decomposition of the edges into simple
cycles. Thus it is bridgeless, and we may color each face by the parity
of the number of these cycles surrounding it. Crossing any edge changes
the color, so the face lengths in each color sum to the number of edges, $14$.
Its nine faces have total length $28$, and every face has length at least
three because the graph is simple and bridgeless. Exactly one face is a
quadrilateral. The color without that face gives $3\mid14$, a contradiction.

Now let $r$ count the threes. If $r\ge5$, the five smallest entries sum
to $15$, below the $16$ required by~Lemma~\ref{sp:lem:eg}. If $r=4$
write $h=(a,b,c,3^4)$, with $c\ge4$,
$a\le b+c-4$ and $a+b+c\ge16$. The graph
$K_2\vee(P_3\sqcup P_2)$ has degrees $(6,6,4,3^4)$, with the first
three vertices forming a triangle. Denote them by $x_0,x_1,x_2$ in
target order. The extra multiplicities on $x_ix_j$ are
\[
 m_{01}=(a+b-c-8)/2,\quad m_{02}=(a+c-b-4)/2,\quad m_{12}=(b+c-a-4)/2.
\]
Their sum is $t$.
If $b\le5$, the two displayed constraints force $(a,b,c)=(6,5,5)$.
For $b\ge6$ the second multiplicity is nonnegative since $a\ge b$
and $c\ge4$. The third is nonnegative since $a\le b+c-4$.
If the first is negative, then $6\le a\le a+b-c<8$. Since $a+b-c$
has the parity of $H$, it must equal six, giving $(6,6,6)$.

To exclude $(6,5^2,3^4)$, remove its universal vertex. An outerplanar
graph of degrees $(4^2,2^4)$ would have eight edges. It must be connected,
since two components of minimum degree two on six vertices can only be
triangles. Deleting a cut vertex leaves five vertices, and every
component has at least two since the original minimum degree is two.
The two components thus have orders two and three. Together with the
cut vertex, they induce outerplanar graphs with at most three and five
edges respectively. Equality forces a triangle glued to a diamond,
giving degrees $(5,3,2^4)$ or $(4,3^2,2^3)$. In the remaining
two-connected case the outer boundary is a hexagon, and both diagonals
would have to join the same two degree-four vertices.

For $(6^3,3^4,2)$, suppress the two-vertex if its neighbors are
nonadjacent. This would give a seven-vertex triangulation with four
degree-three vertices. Such vertices are independent in any triangulation
on more than four vertices (two adjacent ones force all four faces of
a tetrahedron), so their twelve incident edges violate the bipartite
planar edge bound $2\cdot7-4$. If its neighbors are adjacent, remove
the two-vertex to get a seven-vertex plane graph with fourteen edges
and four vertices of degree at most three. None can have degree two,
since deleting it would leave a six-vertex triangulation with at least
three vertices of degree at most three. The only six-vertex triangulations,
the octahedron and $K_2\vee P_4$, have zero and two such vertices,
respectively. Hence both neighbors had
degree six, leaving the previously excluded sequence $(6,5^2,3^4)$.

Finally suppose $1\le r\le3$. Remove a three and reduce a largest entry
by one. The largest entry was at least five, and two entries equal to $U_{\max}$
would leave only sixteen for the other five entries, whose sum is
at least $20-r\ge17$. The residual maximum is at most $U_{\max}-1$, so the
residual is graphic by Lemma~\ref{sp:lem:eg}. Lemma~\ref{sp:lem:insert} lifts any planar
residual. Lowering a maximum of at least five creates no new three,
so the residual has $r-1\le2$ threes. For each remaining exceptional
residual core, the right column lists the
preimages obtained by adjoining a three and raising one of the old entries
by one to a (possibly tied) maximum.
\[
\begin{array}{c|l}
 \text{Residual core}&\text{Possible parent cores}\\
 \hline
 (6,4^5)&(7,4^5,3)\\
 (5,4^4,3)&(6,4^4,3^2),\ (5^2,4^3,3^2)\\
 (5^5,3)&(6,5^4,3^2).
\end{array}
\]
For the parent $(7,4^5,3,2)$, label a hexagon cyclically from $0$ to $5$, add the chords $02,35$, attach a leaf at $1$, and take the cone. Omitting the leaf before coning gives $(6,4^4,3^2)$. Table~\ref{sp:tab:rot-orders} gives the other two.

\paragraph{Seven core vertices on the lower boundary}

If a three occurs, remove one three and lower a largest entry by one.
That entry is at least four, and if it equals $U_{\max}$ it is unique, since
$H-2U_{\max}=14<3(7-2)$. The residual list remains on $\delta=2$ and is graphic
by Lemma~\ref{sp:lem:eg}. The six-core case gives a planar realization
with the same $t$ twos. Reinserting
the three finishes. Otherwise every core degree is at least four and
$t\ge1$. If $a=U_{\max}$, take a hexagon with one chord, then add leaves to
attain the six targets minus one. The number of leaves is
$H-a-20=t$; coning finishes. If $a<U_{\max}$, remove one two. The upper-boundary
residual has maximum at most $U_{\max}-1$ and is graphic by Lemma~\ref{sp:lem:eg}.
We may subdivide one edge of its planar realization unless its core is
$(4^7)$. For that core, subdivide an octahedron edge once and apply the
four-insertion of Lemma~\ref{sp:lem:insert} to the resulting seven-vertex
graph. This proves Proposition~\ref{sp:prop:small}.

\subsection{Eight or more core vertices}\label{sp:sec:large}

\begin{proposition}\label{sp:prop:large}
If $p\ge8$, every leaf-free graphic sequence on either boundary is
planaric.
\end{proposition}

\paragraph{Upper boundary with all core degrees at least five}

\begin{lemma}\label{sp:lem:fan}
Let $a,b,c$ be the three largest entries, in descending order, of a core
$h$ with $p\ge8$ and $h_p\ge5$. Suppose $H$ is even and
$t=H/2-2p\ge0$. If $a\le t+5$ and $a\le U_{\max}=p+t-1$, then $(h,2^t)$
is planaric in either of the following cases.
\begin{enumerate}
\item $b\ge p-1$ and $a+b-c\ge2p-4$.
\item All core entries are distinct.
\end{enumerate}
\end{lemma}

\begin{proof}
Put $a,b$ at the poles of $K_2\vee P_{p-2}$. Their residual demands
are $A=a-p+1$ and $B=b-p+1$. For the second case, strict descent gives
$c\ge p+2$, so both demands are nonnegative there as well.
The path vertices have residual sum $R=H-a-b-4p+10$, and
$R-(A-B)=2(t+5-a)\ge0$. By the double-fan rule it suffices to
place extra path edges leaving capacity at most $A+B$.

Set $q=p-2$ and write $w_1\le\cdots\le w_q=c-4$ for the remaining
target degrees minus four. We may order these capacities along the
path as we please. The two ends each have one further unit of capacity.
For the first assertion, pair consecutive capacities, leaving the
smallest one unpaired when $q$ is odd. In each pair use the smaller
capacity in edge copies. The unused capacities of the pairs are
differences over disjoint intervals in the ordered list, so they sum
to at most $w_q-w_1$. When $q$ is odd the unpaired vertex contributes
a further $w_1$. In either parity at most $w_q$ units remain before
the two endpoint units are counted, and $w_q+2=c-2\le A+B$.

For the second assertion consider four consecutive capacities
$\alpha<\beta<\gamma<\eta$. Pairing them consecutively leaves
$(\beta-\alpha)+(\eta-\gamma)$ unused. Instead order the path as
$\alpha,\gamma,\eta,\beta$ and use $\alpha$ and $\beta$ extra copies
of its first and last edges. This saturates both ends and leaves
$\gamma-\alpha$ and $\eta-\beta$ units at the middle vertices.
Copies of the middle edge can now use one unit from each until only
$|(\eta-\gamma)-(\beta-\alpha)|$ units remain.
Since both outer gaps are positive integers, their difference in absolute value is at most
$(\eta-\gamma-1)+(\beta-\alpha-1)$. Let $s$ be the remainder of $q$
modulo four. Set aside the smallest $s$ entries and apply the block
construction to the rest in increasing order, using zero extra
multiplicity on joins between blocks. For $s=0,1,2,3$ the leftover
small block can have unused capacity respectively
\[
 0,\quad w_1,\quad w_2-w_1,\quad |w_3-w_2-w_1|.
\]
For $s=2$, the difference $w_2-w_1$ is one plus the excess of the
first gap over one. For $s=3$, place $w_3$ in the middle. Its two
incident edges can take $\min\{w_3,w_1+w_2\}$ extra copies in total,
leaving $|w_3-w_1-w_2|$ units. Since both
$w_3-w_2$ and $w_1$ are positive integers, their absolute difference
is at most $(w_3-w_2-1)+(w_1-1)$.
Write $\Delta=w_q-w_1-(q-1)$ for the sum of all gap excesses.
The four-entry blocks and the small block use disjoint gaps. Thus
the unused capacity before counting endpoints is at most $\Delta$,
$w_1+\Delta$, $1+\Delta$, or $w_1-1+\Delta$, respectively.
Since $w_1\ge1$, all four bounds are at most $w_1+\Delta$.
Adding the two endpoint units gives
$w_1+\Delta+2=w_q-q+3=c-p+1\le A\le A+B$.
In either construction the support has $3p-6\ge2p$ edges, so its
cost is at most $t$.
\end{proof}

\begin{lemma}\label{sp:lem:high-core}
Every leaf-free graphic sequence on $S=4n$ with $p\ge6$ and
$h_p\ge5$ has a planar realization.
\end{lemma}
\begin{proof}
Choose a counterexample with $p$ as small as possible. Proposition~\ref{sp:prop:small}
excludes $p=6,7$, so $p\ge8$.
Lemma~\ref{sp:lem:cone} gives $a\le t+4$, and Lemma~\ref{sp:lem:fan} rules
out distinct core degrees. Choose the smallest repeated
value $g$. After removing two copies the core has $p'=p-2\ge6$
vertices and degree sum $H'=H-2g\ge5p'$. To stay on the upper boundary
we take $t'=H'/2-2p'=t-g+4\ge\lceil p'/2\rceil\ge3$.
If the new maximum $a'$ were at most $p'+t'-1=U_{\max}-g+2$, the reduced sequence
would be graphic by Lemma~\ref{sp:lem:eg} and planaric by minimality.
Since $t'\ge2$, Lemma~\ref{sp:lem:pair} would then restore the equal pair.
Hence $U_{\max}-g+2<a'\le a$. Since $U_{\max}=p+t-1$, integrality gives
$a+g\ge p+t+2$. Together with $a\le t+4$, this gives $g\ge p-2$.

We claim $a\ge2p-4$. This holds immediately if $g\ge2p-4$. Otherwise
remove a largest entry and one remaining copy of $g$. Their removal
leaves $p-2$ entries of sum $\sigma=H-a-g$.
Substituting $2t=H-4p$ in $2(a+g)\ge2(p+t+2)$ yields
$a\ge\sigma-g-2p+4$. Since $g\ge p-2\ge6$ is
the smallest repeated value, at most one five occurs, so
$\sigma\ge6(p-2)-1$. Using $g\le2p-5$ gives
\[
 a\ge\sigma-g-2p+4\ge2p-4.
\]
Now $a+b-c\ge a\ge2p-4$, so Lemma~\ref{sp:lem:fan}(1) applies whenever
$b\ge p-1$. Since $b\ge g\ge p-2$, only $b=g=p-2$ remains.
Then $a+g\ge p+t+2$ and $a\le t+4$ force $a=t+4$.

For this last case take $K_2\vee P_{p-2}$ and erase one edge from
the second pole to the path. The second pole now has degree $p-2=b$.
The first pole needs $A=a-p+1$ more incidences, and the path vertices
need $R=H-a-b-(4p-11)$ in total. As $a=t+4$ and $b=p-2$, we have
$R-A=4$. Each path vertex needs at least one incidence, since its
target is at least five and its present degree at most four. Double two
vertex-disjoint path edges and supply every remaining path incidence
on its spoke to the first pole. The support has $3p-7\ge2p$ edges,
so the subdivision count yields a realization with $t$ degree-two vertices.
\end{proof}

\paragraph{Restoring degrees four and three on the upper boundary}

We first prove by induction on $p\ge7$ that every leaf-free graphic
sequence on the upper boundary with minimum core degree at least four
is planaric, with the sole exception of $(4^7)$.
Proposition~\ref{sp:prop:small} gives the base case $p=7$.
For $p\ge8$, Lemma~\ref{sp:lem:high-core} covers minimum core degree at
least five. Suppose then that $h_p=4$. If $a=U_{\max}$, put the
other $q=p-1$ core vertices on a cycle and add two endpoint-disjoint
noncrossing chords. Its maximum degree is three. Attach leaves to
raise the cycle vertices to the corresponding targets minus one.
Since $a=q+t$, precisely $H-a-3q-4=t$ leaves are needed. The cone
is the required simple plane graph. For $a<U_{\max}$, remove a four,
leaving $t$ twos and maximum at most $U_{\max}-1$. The reduced sequence is
graphic by Lemma~\ref{sp:lem:eg}. The residual has $n'\ge p-1\ge7$ vertices
and $m'=2n'$ edges on the upper boundary. Hence $m'<3n'-6$, and
Lemma~\ref{sp:lem:insert} restores the four in any planar realization.
A nonplanaric residual is possible only
for $p=8$ and core $(4^7)$. A square antiprism realizes $(4^8)$.

We now allow threes and induct on $p\ge8$. For a core containing a three,
delete it and reduce the maximum $a$ by one. This leaves the number $t$ of twos unchanged. We have
$a\ge5$ since $H\ge4p$ and the core contains a three.
For $p\ge9$ a universal entry is unique, since two equal to $U_{\max}$ would
leave sum $H-2U_{\max}=2p+2<3(p-2)$. The reduced sequence has maximum at
most $U_{\max}-1$ and at least eight core entries. It is therefore graphic by
Lemma~\ref{sp:lem:eg} and planaric by induction. Lemma~\ref{sp:lem:insert}
restores the three.

At $p=8$ the reduction gives seven core entries and the new degree
bound is $U_{\max}-1$. By Lemma~\ref{sp:lem:eg} the residual is graphic unless
two original entries equal $U_{\max}$. In that case the other six sum to
$H-2U_{\max}=18$, so all six are threes. The original sequence has core
$(a,a,3^6)$ and $t=a-7$. It is realized by
$K_2\vee(3K_2\sqcup tK_1)$, a subgraph of a double fan. If the
residual is graphic and nonplanaric, Proposition~\ref{sp:prop:small}
gives the following possible reduced cores and their preimages.
\[
\begin{array}{c|l}
 \text{Reduced core} & \text{Original core}\\ \hline
 (4^7)&(5,4^6,3)\\
 (6,5^2,3^4)&(7,5^2,3^5),\ (6^2,5,3^5)\\
 (6^3,3^4)&(7,6^2,3^5).
\end{array}
\]
Table~\ref{sp:tab:rot-orders} gives a realization for each of the four preimages
with its required twos. This establishes the induction for the upper boundary.

\paragraph{The lower boundary}

Here $t=H/2-2p+1$. When a three occurs, delete it and lower a largest
entry by one. We have $a\ge4$. If $a=U_{\max}$ then it is unique since
$H-2U_{\max}=2p<3(p-2)$. The reduced sequence still has $t$ twos, its
maximum is at most $U_{\max}-1$, and it is graphic by Lemma~\ref{sp:lem:eg}
since $p-1\ge6$. Induction on $p$ using Proposition~\ref{sp:prop:small} and
Lemma~\ref{sp:lem:insert} thus handles this case.

If all core entries are at least four, then $t\ge1$. When $a=U_{\max}$ use
a cycle on $q=p-1$ with one chord, complete its degrees to the targets
minus one by adding $H-a-3q-2=t$ leaves, and cone. When $a<U_{\max}$,
remove a single two. The resulting upper-boundary sequence has $t-1$
twos and maximum at most $U_{\max}-1$, so it is graphic by Lemma~\ref{sp:lem:eg}.
The upper-boundary result just proved realizes it
for $p\ge8$. Subdividing any edge restores the lost two and finishes
the proof of Proposition~\ref{sp:prop:large}.

\subsection{Restoring the leaves}\label{sp:sec:leaves}

\begin{lemma}\label{sp:lem:leaf}
Suppose a positive graphic sequence contains a one and its maximum entry
$D$ is at least three. Delete a one, lower a maximum entry by one and reorder.
The resulting positive sequence is graphic. If it is planaric, so is the
original sequence.
\end{lemma}
\begin{proof}
In a realization let the chosen leaf $z$ have neighbor $u$, and choose
a vertex $v$ of degree $D$. If $u\ne v$, then $\deg(v)\ge\deg(u)$
and $|N(v)\setminus\{u\}|>|N(u)\setminus\{v,z\}|$. There is thus
a $w\in N(v)\setminus\{u\}$ adjacent to neither $u$ nor $z$, with
$w\ne z$. Replace $zu,vw$ by $zv,uw$. The degrees are unchanged and
$z$ can now be deleted. In the reverse construction a leaf can always
be attached to the prescribed vertex in a planar embedding.
\end{proof}

\begin{proof}[Proof of Theorem~\ref{sp:thm:first}]
For $\delta\ge4$ the conclusion is due to Bar-Noy et al.~\cite[Theorem~2]{bn}.
By parity it remains to consider $\delta=2$. We induct on $\ell$.
For $\ell=0$, Propositions~\ref{sp:prop:small} and~\ref{sp:prop:large} give
the unique exception $(4^5,2)$. Take $\ell>0$. If the maximum degree is
at most two, every realization is planar. Otherwise Lemma~\ref{sp:lem:leaf}
gives a graphic residual with $\ell-1$ leaves and the same $\delta$,
since $(n,S,\ell)$ becomes $(n-1,S-2,\ell-1)$. By induction we can attach
the deleted leaf unless the residual is $(4^5,2)$. The only entry we
could raise to recover the deleted maximum is a four, so the original
list is $(5,4^4,2,1)$. To realize it,
delete an edge $ab$ from $K_4$ and draw the resulting graph with all
vertices on the outer face. Attach a leaf $x$ to $a$ in that face, cone
the resulting outerplane graph, and then attach a new leaf to $b$. The cone vertex has
degree five, the four original vertices have degree four and $x$ has
degree two, as required.
\end{proof}

\begin{proof}[Proof of Theorem~\ref{sp:thm:next}]
We induct on $\ell$. Propositions~\ref{sp:prop:small} and~\ref{sp:prop:large}
give the base case $\ell=0$, and the core cut count settles the first two
classes of Theorem~\ref{sp:thm:next} for every $\ell$. Outside these classes
take $\ell>0$. Since $H=4p+2t+\ell>4p$, a maximum entry is at least
five. Lemma~\ref{sp:lem:leaf} thus preserves $\delta=0$, $p$, $t$ and the
number of threes. If the residual is planaric, attaching the leaf
finishes the construction. By induction we need only consider residuals
in the last two classes of the theorem.

When $\ell=1$ there are six such residuals. In reversing the deletion one may raise
any entry that, after its increment, is a maximum. Table~\ref{sp:tab:leaf}
lists all such choices. The middle column gives the degrees before the
leaf was deleted, with that single leaf omitted. Each sequence in the last
column is obtained by lowering a different vertex in the middle column.
Attaching a leaf at the vertex just lowered in a planar realization
of the alternative residual gives a planar realization of the predecessor.

\par\medskip\noindent\begin{minipage}{\linewidth}
\centering
\tablecaption{One-leaf predecessors of the six remaining leaf-free exceptions. In the three universal-vertex cases, deleting the leaf recovers the obstruction in the first column.}\label{sp:tab:leaf}
\begin{tabular}{@{}lll@{}}
\toprule
Leaf-free obstruction & Predecessor (leaf omitted) & Residual or forced neighbour \\
\midrule
$(5,4^4,3)$ & $(6,4^4,3)$ & universal vertex \\
            & $(5^2,4^3,3)$ & $(5^2,4^2,3^2)$ \\
$(5^5,3,2^2)$ & $(6,5^4,3,2^2)$ & $(6,5^3,4,3,2^2)$ \\
$(6,4^5,2)$ & $(7,4^5,2)$ & universal vertex \\
$(4^7)$ & $(5,4^6)$ & $(5,4^5,3)$ \\
$(6,5^2,3^4)$ & $(7,5^2,3^4)$ & universal vertex \\
            & $(6^2,5,3^4)$ & $(6^2,4,3^4)$ \\
$(6^3,3^4,2)$ & $(7,6^2,3^4,2)$ & $(7,6,5,3^4,2)$ \\
\bottomrule
\end{tabular}
\end{minipage}\par\medskip

The first alternative residual is realized by the double fan $K_2\vee P_4$.
The other four have maximum degrees $6,5,6,7$ in table order and
respective degree bounds $U_{\max}=7,6,6,7$. Each is therefore graphic by
Lemma~\ref{sp:lem:eg} and planaric by the leaf-free
classification. Attaching the leaf in each case realizes its predecessor.

The other three predecessors begin the families in part~3. For
$\ell\ge2$ the residual still has a leaf, so by induction it belongs
to one of these families. Its universal vertex exceeds every other
degree by at least two. Only that entry can be raised to recover a
maximum, giving the next member of the same family.
Conversely, every member is graphic by attaching leaves to the maximum
of its leaf-free endpoint. In every realization the maximum is universal,
so deleting its leaves recovers that endpoint. Its nonplanarity rules
out a planar realization of the entire family.
\end{proof}

\noindent\begin{minipage}{\linewidth}
\centering
\tablecaption{Explicit planar realizations of the seven lists left by the reductions.}\label{sp:tab:rot-orders}
\small
\setlength{\tabcolsep}{3pt}
\begin{tabular}{@{}l*{9}{>{\ttfamily}c}r@{}}
\toprule
 & \multicolumn{9}{c}{Vertex $v$} & \\
\cmidrule(lr){2-10}
Degree sequence & $0$ & $1$ & $2$ & $3$ & $4$ & $5$ & $6$ & $7$ & $8$ & Faces\\
\midrule
$(6,4^6,2)$ & 4576 & 2653 & 4613 & 1542 & 0623 & 3160 & 512407 & 60 &  & 10\\
$(6,5^4,3^2,2^2)$ & 162435 & 05326 & 14706 & 41508 & 23807 & 310 & 201 & 42 & 34 & 11\\
$(5^2,4^3,3^2)$ & 16352 & 02436 & 1054 & 4061 & 2531 & 420 & 301 &  &  & 9\\
$(5,4^6,3)$ & 36754 & 2576 & 3416 & 0426 & 5230 & 1407 & 1032 & 501 &  & 10\\
$(7,5^2,3^5)$ & 1543762 & 02345 & 10673 & 201 & 105 & 401 & 207 & 602 &  & 10\\
$(6^2,5,3^5)$ & 165724 & 032756 & 13407 & 214 & 302 & 106 & 501 & 201 &  & 10\\
$(7,6^2,3^5,2)$ & 1764325 & 052678 & 150346 & 204 & 302 & 210 & 201 & 108 & 71 & 11\\
\bottomrule
\end{tabular}

\par\smallskip
\raggedright
The digits in column $v$ list its neighbours in cyclic order. Vertex labels need not follow the sorted degree order. To trace a face, follow $(u,v)$ by $(v,w)$,
where $w$ follows $u$ in the cyclic list at $v$. The number $f$ of
cycles of this permutation appears in the last column. Since
$n-m+f=2$ in every row, capping the face walks with disks gives a
sphere embedding.
\end{minipage}

\section{Finding a realization by reading triangular faces}\label{sec:search}
\subsection{The states}
For $n\le3$ enumerate the simple graphs. For $n\ge4$ reject entries above $n-1$ and sums odd or above $6n-12$. Any planar realization extends to an edge-maximal simple plane graph $T$ on the same vertices, including isolates. Maximality rules out distinct components on a common face and consecutive edges at a cut vertex leading to different components of its deletion. Every face boundary is therefore a cycle. Among four consecutive corners, the two diagonals between alternate corners cannot both lie outside the face, so every face is triangular. Colour the original edges red and the added edges blue. The search tries possible faces and edge colours one at a time. A complete run reads $f=2n-4$ faces and counts each red edge at its first appearance.

An edge used by one face so far is \emph{open}, and an edge used twice is \emph{sealed}. The active vertices are the endpoints of the open edges. A state consists of the following data.
\begin{enumerate}
\item The numbers of faces read and vertices introduced, together with the multiset of target degrees still to be assigned. We remove a vertex's degree from this multiset only when the vertex ceases to be active.
\item All open and sealed edges between active vertices. We retain a sealed edge to prevent a later face from creating a second edge with the same endpoints. The open edges are partitioned into blocks according to which component of the read-face dual contains their incident read face.
\item The red degree accumulated at each active vertex. A red edge contributes at its first appearance.
\end{enumerate}
We identify states under bijections of the active vertices preserving these data. The colour of a newly seen edge affects only its two endpoint counts, so the state need not store it separately.

To add a face, choose three distinct vertices, each active or previously unused. The choice is admissible only if no two are joined by a sealed edge. Any open side of the new face becomes sealed, while each previously absent side becomes an open edge and receives its colour. In the read-face dual the new face joins all the blocks it meets across old open edges, so we merge these blocks and include its new open sides in the resulting block. If it meets no block, it starts one. Once a block loses its last open edge, no further face can attach to it. We therefore allow this to happen only for the sole remaining block on the final step.

\begin{figure}[htbp]\centering
\begin{tikzpicture}[scale=.9,completion/.style={draw=blue!65!black,thin}]
\begin{scope}
\coordinate (a) at (-1,0);\coordinate (b) at (1,0);\coordinate (c) at (0,1.35);
\fill[accent!8] (a)--(b)--(c)--cycle;
\draw[keep] (a)--(b)--(c);\draw[completion] (c)--(a);
\foreach \i/\pos in {a/left,b/right,c/above}\node[v,label=\pos:{$\i$}] at (\i){};
\node at (0,-2.12){(a) $abc$};
\end{scope}
\begin{scope}[xshift=5.2cm]
\coordinate (a) at (-1,0);\coordinate (b) at (1,0);\coordinate (c) at (0,1.35);\coordinate (d) at (0,-1.35);
\fill[accent!8] (a)--(c)--(b)--(d)--cycle;
\draw[keep] (c)--(b) (a)--(d);\draw[completion] (a)--(c) (d)--(b);\draw[keep,double=white,double distance=.8pt] (a)--(b);
\foreach \i/\pos in {a/left,b/right,c/above,d/right}\node[v,label=\pos:{$\i$}] at (\i){};
\node at (0,-2.12){(b) $abc,abd$};
\end{scope}
\end{tikzpicture}
\caption{Two successive faces of a tetrahedron. Thick red edges belong to the realization and thin blue edges complete the host. Reading $abd$ seals $ab$ (double line). Only $ad$ adds to the red degrees, now $(2,2,1,1)$ at $(a,b,c,d)$. The four outer edges form one block of open edges.}\label{fig:front}
\end{figure}
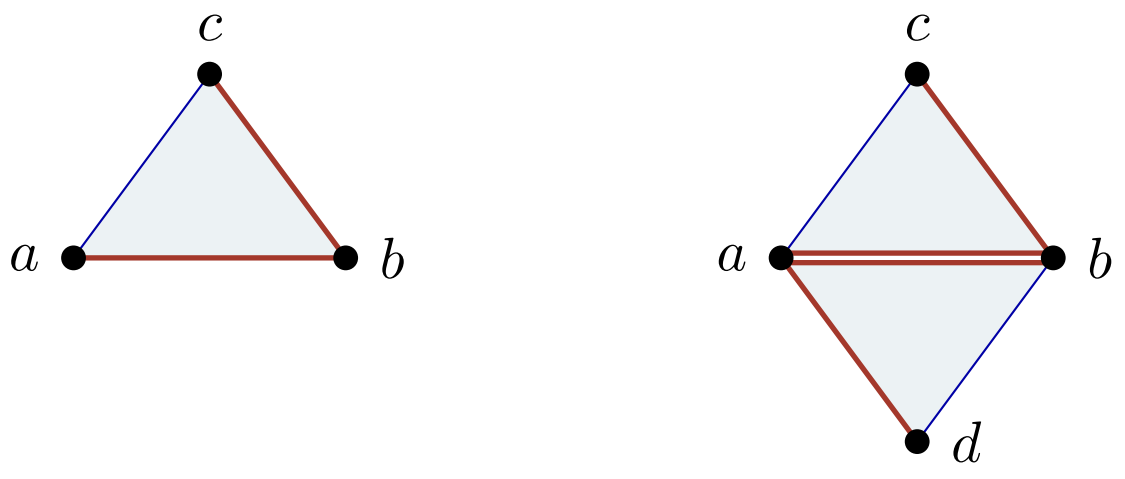

When a vertex loses its last open edge, its red degree is final. We remove one copy of that degree from the residual multiset, rejecting the step if none remains. We can now forget all sealed edges incident with the vertex, since a retired name is never used again. The search starts with an empty boundary, the entire input multiset and $n$ unused names. After $f$ steps it accepts if the sole face component has closed, all $n$ names have been used and the residual multiset is empty. Figure~\ref{fig:front} illustrates the transition across a shared edge.

\subsection{What an accepting run represents}
The states forget the cyclic order at each vertex. To recover a sphere at acceptance, we must rule out several disjoint fans of triangles meeting at a vertex.
\begin{lemma}\label{lem:faces}
For any sphere triangulation whose red subgraph has the input degrees, the unrestricted transitions accept every ordering of its faces. Conversely, every accepting run gives a simple planar realization of the input.
\end{lemma}
\begin{proof}
In a triangulation the faces around a vertex form a cycle. If some of them have been read and others have not, the cycle crosses from a read face to an unread face across an incident open edge. Consequently a vertex needed later is still available. The full face dual is connected, so every proper component of read faces has an open edge. A true face ordering obeys all transition and retirement rules.

Conversely, follow an accepting run, assigning a different permanent name to every introduced vertex. A pair of active vertices cannot reopen a sealed edge, and a pair with a retired endpoint cannot occur again. Each edge thus has exactly two occurrences at closure. The transition tests do allow two face instances with the same unordered triple of corners. Those two faces would use up both occurrences of all three sides and hence form an entire component of the face dual. At acceptance the dual is connected and $f=2n-4\ge4$, so no triple is repeated.

Glue the triangles along paired sides, matching their named endpoints. The interior of every edge has a disk neighbourhood. To examine a vertex, consider its link, with one vertex for each incident edge and one edge for each face corner. Since every side occurs twice, every vertex of the link has degree two. The link is therefore a disjoint union of circles. Splitting the original vertex into one copy per circle gives each copy a disk neighbourhood, and faces sharing an edge still share it after splitting. The connected face dual thus gives a connected closed surface.

If splitting adds $h$ vertices, the surface has $n+h$ vertices, $3f/2$ edges and $f$ faces. Its Euler characteristic is
\[
 n+h-3f/2+f=2+h.
\]
By the classification of compact surfaces, a connected closed surface has Euler characteristic at most two, and equality characterizes the sphere, with or without an orientability assumption. Hence $h=0$ and the gluing itself lies on the sphere. Distinct corners and the pair rule give a simple graph. Its red subgraph realizes the input because each red edge was counted once at each endpoint and every vertex retired exactly once.
\end{proof}

\subsection{A boundary catalogue independent of the host}
A given embedding supplies boundaries for a surface decomposition~\cite{rst}. For triangulating a prescribed point set, Lingas~\cite{lingas} and Marx and Miltzow~\cite{mm} choose separators from the given points. For degree realization the catalogue must be built from $n$ alone.

\begin{lemma}[Boundary catalogue]\label{lem:catalogue}
For every $n\ge4$ there is a deterministic $2^{O(\sqrt n)}$-time algorithm that, given $n$, lists at most $2^{O(\sqrt n)}$ records of $O(\sqrt n)$ bits each. For every red--blue sphere triangulation on $n$ vertices, its faces admit an order in which each prefix has a matching record. A match is a bijection from the active vertices of the prefix to those of the record that preserves the current red degrees, the open and retained sealed edge sets, and the partition of the open edges into read-face components.
\end{lemma}

\begin{proof}
\emph{Choosing the order.}
An unweighted separator bounds the number of vertices we must remember. Charging each of them $\lceil\log_2 n\rceil$ bits would still leave a logarithmic factor in the exponent. We charge vertex $v$ of the host $T$ only $c(v)=\lceil\log_2(2+\deg_T(v))\rceil$, enough to store any red degree it can attain, and give it weight $w(v)=c(v)^2$. Since $c(v)^2\le9(\deg_T(v)+1)$ and $\sum_v\deg_T(v)=6n-12$, we have $w(T)\le63n$.

The vertex separator theorem of Aleksandrov et al.~\cite[Theorem~2]{sep} applies to any planar graph $H$ with nonnegative vertex costs and weights. For $0<t<1$ it gives a set of cost at most $4\sqrt{2\sum_{v\in H}c(v)^2/t}$ whose removal leaves each component with weight at most $t\,w(H)$. With the weights above and $t=1/2$, this costs at most $8\sqrt{w(H)}$. Recurse on the remaining components, assigning every removed vertex to the node where it is removed. The weights are positive and halve at each step. The sum of the separator costs along a root-to-leaf path is therefore bounded by
\begin{equation}\label{eq:cost}
 8\sqrt{w(T)}\sum_{i\ge0}2^{-i/2}=O(\sqrt n).
\end{equation}
Vertices assigned to incomparable nodes lie in different components after their last common separator is removed. Hence the endpoints of an edge belong to comparable nodes, and the three nodes containing the corners of a face lie on one path. Assign each face to the deepest of these nodes and read the faces at each node before visiting its children in depth-first order. While reading the faces at $z$, every vertex in a completed side branch has seen all its incident faces, and every vertex in an untouched subtree has seen none. The active vertices therefore lie on the separators along the path to $z$. If there are $b$ such vertices and $r_v$ is the current red degree of $v$, then $r_v\le\deg_T(v)$ gives
\[
 b\le\sum_{v\text{ active}}\lceil\log_2(2+r_v)\rceil=O(\sqrt n).
\]

\smallskip\noindent\emph{Encoding the boundary.}
Let $J$ consist of the retained edges on the $b$ active vertices, with $s=|E(J)|\le3b$. Subdivide its edges at their midpoints. To encode the open-edge partition in the same plane drawing, cut the read-face dual wherever it crosses $J$. Each cut-dual component now joins read face centres only through unretained sides, so it can be contracted disjointly from $J$. Keep the components incident with $J$ and draw a spoke from each retained side's midpoint to the component of each incident read face. An open midpoint has one spoke and a sealed midpoint has two, with duplicate spokes suppressed. Passing through a sealed midpoint restores exactly the dual adjacency cut there. Thus two open midpoints are joined by a path of spokes precisely when their read faces were connected. The marked plane graph has $2s$ primal segments, at most $2s$ spokes and $O(b)$ vertices.

As long as the drawing is disconnected, some face has vertices from two components on its boundary. Join such a pair by an arc inside the face and mark the new edge for deletion. Each step reduces the component count by one and preserves planarity. There are at most $b+3s$ vertices, so the connected drawing still has only $e=O(b)$ edges. Choose a spanning tree and a root corner. A clockwise tour of the tree records an opening and a closing symbol for each tree edge, with a placeholder at every non-tree half-edge encountered on the tour. Cutting the sphere along the tree leaves a disk whose boundary follows this tour. The non-tree edges run through that disk as disjoint arcs, so their endpoints form a noncrossing matching of the placeholders. Writing an opening symbol at the first endpoint of each arc and a closing symbol at the second uniquely records the matching, since each closing symbol pairs with the most recent unmatched opening. Together the two words specify the tree, the corner of every placeholder and the edge incident there.

Prefix each word of $\ell$ symbols with $1^\ell0$. The rooted tour numbers vertices and edges by their first occurrence, with the root vertex first. In this order, append two bits per vertex for its type (original, open midpoint, sealed midpoint or dual component) and two per edge for its type (primal segment, spoke or bridge). The two words and the marks take $O(e+1)$ bits. Finally, for each active vertex append the $k_v$ binary digits of $r_v+1$ preceded by $1^{k_v}0$, where $k_v=\lceil\log_2(2+r_v)\rceil$. These fields take at most $3\sum_v k_v$ bits. One initial bit distinguishes the empty drawing. The size bound on $e$ and~\eqref{eq:cost} give a fixed constant $C$ for which the whole record fits in $C\lceil\sqrt n\rceil$ bits.

Enumerating all strings up to this length gives the catalogue. A string must decode in full into the two words, their tree traversal and matching, and the required marks and counters. After the bridges are deleted, every primal segment must join an original vertex to a midpoint. Exactly two such segments must meet each midpoint, with distinct original endpoints, and no two midpoints may give the same endpoint pair. Every original vertex must meet an open midpoint. Spokes join midpoints to component vertices, with one spoke at an open midpoint, one or two at a sealed midpoint, and at least one at each component vertex. The open midpoints in each connected component of the spoke subgraph form a block whenever this set is nonempty. We require at most $n$ original vertices and one counter in $[0,n-1]$ for each. These tests take polynomial time per string. The surviving records, including the empty boundary, form the \emph{allowed table} in any fixed order. Several records may represent the same boundary, and some may occur in no triangulation. The construction above ensures that every prefix of the chosen face order has a record in the table.
\end{proof}

\paragraph{Using the catalogue.}
Assign each proposed successor the index of its first isomorphic record, discarding it if none matches. Suppose two prefixes agree on the counts and residual multiset, and their active data are isomorphic. For any next face, the isomorphism identifies its active corners. Equal numbers of unused names let us pair the fresh corners and give the new edges the same colours. Both steps then merge corresponding blocks, remove the same degrees and again have isomorphic active data. Induction shows that the prefixes admit the same continuations, so one history per state suffices. Each retained history consists of actual transitions, and Lemma~\ref{lem:faces} applies if it accepts.

\subsection{Counting states and recovering the graph}
The data to be compared have a natural representation by a coloured incidence graph. Its four colour classes distinguish active vertices, open edges, sealed edges and blocks, with the first class further refined by current red degree. Each edge vertex is adjacent to its two endpoints, and each open-edge vertex also to its block. Colour-preserving isomorphism of these graphs is precisely equivalence of the active data. The plane drawing also gives a short path decomposition for this comparison.

Let $K=4C\lceil\sqrt n\rceil$, an upper bound on the sizes of the decoded drawings and incidence graphs. For a true boundary, apply the separator theorem to its plane drawing with unit costs and weights. Visit the recursion tree in depth-first order, with a bag at each node containing its separator and all ancestor separators. A vertex occurs precisely during the visit to its subtree, and the ends of every edge share a bag. The geometric sum in~\eqref{eq:cost} now bounds the bag size by $O(\sqrt K)$.

To pass to the incidence graph, delete the temporary bridges and split each midpoint into two copies in every bag containing it. Give one copy the two primal segments and the other its spokes. Link the copies of an open midpoint, whose edge needs a block incidence. Leave the copies of a sealed midpoint unlinked. Contract each connected component of the spoke subgraph to one vertex, keeping it as a block only if it contains an open midpoint. The bags meeting a connected subgraph form an interval, so the contraction preserves the path decomposition. The splitting at most doubles the bag size. Write $w$ for twice the ceiling of $8\sqrt K/(1-1/\sqrt2)$, an upper bound on the resulting bag size.

Order the vertices by their last bag, breaking ties arbitrarily. At the end of any nonempty prefix, every unprocessed neighbour of a processed vertex belongs to the last bag of the most recently processed vertex. Thus the prefix has at most $w$ external neighbours. We can find an order with this property by a subset search, extending a reached set by one vertex whenever the enlarged set has at most $w$ external neighbours. Keeping one order per reached set costs $2^{O(K)}$ time per pattern. We discard patterns with more than $K$ vertices or with no such order. Every boundary required for completeness survives these tests.

To compare a surviving pattern $P$ with a candidate $H$, first check that their orders and edge counts agree. For each $X\subseteq V(H)$ count the colour-preserving homomorphisms $P\to H[X]$. In the saved order $v_1,\ldots,v_k$ of $P$, let $S_i=\{v_1,\ldots,v_i\}$ and let $B_i=N(S_i)\setminus S_i$ be its external neighbours. For each assignment on $B_i$, store the number of extensions to $S_i$ preserving the colours there and the edges incident with $S_i$. Since $B_{i-1}\setminus\{v_i\}\subseteq B_i$, an assignment on $B_i$ and a choice for the image of $v_i$ supply the entire preceding table key. Check the colour of $v_i$ and its edges to later neighbours, and sum over its possible images. This costs $K^{O(w+1)}$ time for each $X$.

Inclusion--exclusion over $X$ counts the maps onto $V(H)$. With equal orders these maps are bijections, and with equal edge counts they are isomorphisms. Even the sum of the absolute values in the inclusion--exclusion calculation is at most $2^K K^K$ for $K\ge1$. We thus need only $O(K\log(K+2))$ bits per integer. Since $K=O(\sqrt n)$ and $w=O(\sqrt K)=O(n^{1/4})$, the bag factor is $2^{O(n^{1/4}\log n)}=2^{o(\sqrt n)}$. A comparison, successful or otherwise, takes $Q(n)=2^{O(K)}K^{O(w+1)}=2^{O(\sqrt n)}$ time.

The positive entries of the residual multiset form a partition of an integer at most $6n$, while the number of zeros can take at most $n+1$ values. Write $p(m)$ for the ordinary partition number. The Hardy--Ramanujan asymptotic formula~\cite{hr} gives $p(m)=\exp(O(\sqrt m))$, and hence $P(6n)=\sum_{m=0}^{6n}p(m)=2^{O(\sqrt n)}$. If $A$ is the allowed-table length, the number $S$ of possible state keys is at most $(f+1)(n+1)^2 P(6n)A=2^{O(\sqrt n)}$, including the introduced-vertex count and the multiplicity of zero.

There are $f+1$ layers indexed by the number of faces read and at most $S$ keys per layer. Keep one history with permanent labels in $[n]$ per key. Each has at most $8n^3$ extensions, one for each choice of corners and edge colours. Finding the first matching catalogue record costs at most $AQ(n)$ per extension. Its index, the global counts and the residual multiset identify the successor. Scanning the next layer for this key costs $S n^{O(1)}$, since a history has at most $2n-4$ faces and takes polynomial time to copy or update. Together these costs give $2^{O(\sqrt n)}$ time. An accepting history records every red edge at its first appearance. Matching final degrees to equal input entries recovers the edges on the given labels.

On a deterministic multitape Turing machine, a first scan tallies the entries to obtain $n$. A second scan skips leading zeros, records all-zero entries as zero, and rejects a positive entry as soon as its significant part exceeds $n-1$. Only $O(\log(n+1))$ significant bits per entry need examination or storage. The scans take $O(L)+n^{O(1)}$ steps. All subsequent tables and histories together use $2^{O(\sqrt n)}$ bits, so even a full tape scan per table access multiplies the search time by at most $2^{O(\sqrt n)}$. This proves Theorem~\ref{th:alg}.

\begin{corollary}\label{cor:compact-listing}
The potentially planar degree multisets on $n$ vertices can be listed without repetition in $2^{O(\sqrt n)}$ total time. Their number is $2^{\Theta(\sqrt n)}$.
\end{corollary}
\begin{proof}
Generate the partitions of integers at most $6n$ with at most $n$ parts by appending parts in nonincreasing order. Each prefix is itself a partition in the same range. Pad with zeros and apply Theorem~\ref{th:alg} to each of the $2^{O(\sqrt n)}$ candidates. For $n\ge2$, trees alone give $p(n-2)$ different multisets. Pad a partition of $n-2$ to length $n$ and take the resulting entries as the label multiplicities in a Pr\"ufer word. The tree degrees recover those entries by subtracting one. The partition asymptotic~\cite{hr} gives the claim.
\end{proof}

\paragraph{Data and acknowledgements.}
The Lean~4 companion~\cite{companion} indexes the formal results by theorem number and includes scripts for the finite enumerations. The interval rule, the search algorithm and the arithmetic criterion use the canonical-ordering theorem, the cost-weight planar separator theorem and the weighted matrix-tree theorem, respectively, as explicit hypotheses in Lean. We thank the members of The Kai Project for discussions on AI-assisted verification.
\paragraph{Declaration of generative AI use.}
The authors directed the research, developed the mathematical arguments, and prepared and revised the manuscript. OpenAI Codex (including GPT-6.2 Astra) was used to explore and check candidate proofs, run computational checks, search the literature, draft and edit text, and write Lean, Python and TikZ code. The authors take full responsibility for the manuscript.

\end{document}